\documentclass[12pt]{article}
\usepackage{graphicx,amssymb,amsmath,empheq,color,comment}
\usepackage{amsthm}
\usepackage[colorlinks=true,urlcolor=blue,linkcolor=black,citecolor=black]{hyperref}
\usepackage{subfig} 

\makeatletter

\usepackage{tikz}
\usepackage{tikz-cd}
\usetikzlibrary{arrows}
\usetikzlibrary{intersections}
\usetikzlibrary{shapes.geometric}
\usetikzlibrary{decorations.pathmorphing, decorations.markings, patterns,shapes}

\tikzset{
  baseline = -0.5ex,
  wavy/.style = {
    thick,
    decorate,
    decoration={snake,amplitude=2pt,segment length=5pt}},
  sdot/.style = {
    circle,
    draw=none,
    fill=black,
    minimum size=2.5pt,
    inner sep=0pt},
  bdot/.style = {
    circle,
    draw=none,
    fill=black,
    minimum size=4pt,
    inner sep=0pt},
  svertex/.style = {
    circle,
    draw=black,
    thick,
    fill=lightgray,
    minimum size=8pt,
    inner sep=1pt},
  mvertex/.style = {
    circle,
    draw=black,
    thick,
    fill=lightgray,
    minimum size=12pt,
    inner sep=1pt},
  bvertex/.style = {
    circle,
    draw=black,
    thick,
    fill=lightgray,
    minimum size=16pt}}

\tikzset{
  mid arrow/.style={postaction={decorate,decoration={
        markings,
        mark=at position .575 with {\arrow[#1]{stealth}}
      }}},
  near arrow/.style={postaction={decorate,decoration={
        markings,
        mark=at position .275 with {\arrow[#1]{stealth}}
      }}},
   far arrow/.style={postaction={decorate,decoration={
        markings,
        mark=at position .800 with {\arrow[#1]{stealth}}
      }}},
}

\newlength{\fighskip} \fighskip=2pt
\newlength{\figvskip} \figvskip=3pt

\newcommand*{\widebox}[1]{\setlength{\fboxsep}{1ex}%
  \fbox{#1}}
\newcommand*{\wideboxed}[1]{\setlength{\fboxsep}{1ex}%
  \fbox{\m@th$\displaystyle#1$}}

\def\ubrace#1_#2{%
  \underbrace{#1}_{\hb@xt@\z@{\hss$\scriptstyle#2$\hss}}}

\newcommand{\blangle}{\bigl\langle}
\newcommand{\brangle}{\bigr\rangle}
\newcommand{\dlangle}{\langle\kern-1.5pt\langle}
\newcommand{\drangle}{\rangle\kern-1.5pt\rangle}
\newcommand{\bdlangle}{\blangle\kern-3pt\blangle}
\newcommand{\bdrangle}{\brangle\kern-3pt\brangle}

\newcommand*{\bra}[1]{\langle#1|}
\newcommand*{\ket}[1]{|#1\rangle}
\newcommand*{\braket}[2]{\langle#1|#2\rangle}

\newcommand*{\bbra}[1]{\blangle#1\big|}
\newcommand*{\bket}[1]{\big|#1\brangle}
\newcommand*{\bbraket}[2]{\blangle#1\big|#2\brangle}

\newcommand*{\corr}[1]{\langle{#1}\rangle}
\newcommand*{\bcorr}[1]{\blangle{#1}\brangle}

\newcommand{\vr}{\varrho}
\newcommand{\kap}{\varkappa}
\newcommand{\vth}{\vartheta}

\newcommand{\calH}{\mathcal{H}}

\newcommand{\calJ}{\mathcal{J}}

\newcommand{\calL}{\mathcal{L}}

\newcommand{\calO}{\mathcal{O}}

\newcommand{\ZZ}{\mathbb{Z}}

\newcommand{\RR}{\mathbb{R}}
\newcommand{\CC}{\mathbb{C}}
\newcommand{\LL}{\mathbb{L}}

\DeclareMathOperator{\Tr}{Tr}

\DeclareMathOperator{\TT}{\mathbf{T}}

\newcommand{\const}{\mathrm{const}}

\let\Re\relax\DeclareMathOperator{\Re}{Re}
\let\Im\relax\DeclareMathOperator{\Im}{Im}

\DeclareMathOperator{\OTOC}{OTOC}
\newcommand{\R}{\mathrm{R}}
\newcommand{\A}{\mathrm{A}}
\newcommand{\K}{\mathrm{K}}
\newcommand{\W}{\mathrm{W}}
\newcommand{\gr}{\mathrm{gr}}
\newcommand{\VF}{\Upsilon}
\newcommand{\tVF}{\widetilde{\Upsilon}}
\newcommand{\bdot}{\boldsymbol{\cdot}}

\newcommand{\ta}{\tilde{a}}
\newcommand{\tchi}{\tilde{\chi}}
\newcommand{\tpsi}{\tilde{\psi}}
\newcommand{\tXi}{\widetilde{\Xi}}

\renewcommand{\leq}{\leqslant}
\renewcommand{\geq}{\geqslant}

\makeatother

\numberwithin{equation}{section}

\title{A two-way approach to out-of-time-order correlators}

\author{Yingfei Gu, Alexei Kitaev, Pengfei Zhang\\
\normalsize\it California Institute of Technology, Pasadena, CA 91125, U.S.A.
\vspace{0.5cm}}

\date{March 3, 2022}

\begin{document}

\setcounter{tocdepth}{2}

\maketitle

\begin{abstract}
Out-of-time-order correlators (OTOCs) are a standard measure of quantum chaos. Of the four operators involved, one pair may be regarded as a source and the other as a probe. A usual approach, applicable to large-$N$ systems such as the SYK model, is to replace the actual source with some mean-field perturbation and solve for the probe correlation function on the double Keldysh contour. We show how to obtain the OTOC by combining two such solutions for perturbations propagating forward and backward in time. These dynamical perturbations, or scrambling modes, are considered on the thermofield double background and decomposed into a coherent and an incoherent part. For the large-$q$ SYK, we obtain the OTOC in a closed form. We also prove a previously conjectured relation between the Lyapunov exponent and high-frequency behavior of the spectral function.
\end{abstract}

\tableofcontents
\newpage

\section{Introduction}

Out-of-time-order correlators (OTOCs) are interesting objects for several reasons. First, they characterize quantum chaos in a way that is comparable to the classical picture of divergent trajectories~\cite{LaOv69}. For systems in which connected correlators are suppressed by a large factor $C$, OTOCs behave as $1-C^{-1}e^{\kap t}$ as long as $C^{-1}e^{\kap t}\ll 1$; at later times, they decay to zero. The number $\kap$, called the Lyapunov exponent, is bounded by $\frac{2\pi}{\beta}$~\cite{MSS15}. OTOCs give access to the near-horizon region of black holes~\cite{ShSt13,Kit.BPS,ShSt14}, in which case $\kap=\frac{2\pi}{\beta}$. Another reason to study OTOCs is their relevance to quantum scrambling and certain information-theoretic tasks~\cite{HaPr07,HQRY15,YoKi17}.

In this paper, we give a conceptual picture and a general expression for OTOCs in many-body systems with all-to-all interactions such as the SYK model~\cite{SaYe93,Kit.KITP.1,Kit.KITP.2,KS17-soft}. For such systems, the previously mentioned parameter $C$ is proportional to the number of elementary degrees of freedom $N$. The early-time behavior is already well-studied. For the SYK model at low temperature, the Lyapunov exponent $\kap$ matches that of a black hole~\cite{Kit.KITP.1}. Maldacena and Stanford~\cite{MS16-remarks} found a finite-temperature correction to $\kap$ and calculated the Lyapunov exponent at an arbitrary temperature in the large-$q$ limit. A general ansatz for the early-time OTOC was introduced in~\cite{KS17-soft}, and a relation between the ratio $C/N$ and the Lyapunov exponent was established in~\cite{GuKi18}.

In principle, the calculation of a correlator like $\corr{(A(t)B(0)C(t)D(0)}$ can be carried out by propagating $B(0)$, $D(0)$ forward or $A(t)$, $C(t)$ backward in time using the Heisenberg equation. This leads to the operator growth picture~\cite{RSS14}. At infinite temperature, the decomposition of the Heisenberg-evolved operator $\calO$ into basis operators (i.e.\ products of $\chi_j$ in the SYK case) may be interpreted as a quantum wave function~\cite{MSS18}. Qi and Streicher~\cite{QiSt18} have extended this method to finite temperatures by constructing the quantum state $\ket{\Psi}=\bket{\calO\rho^{1/2}}$, where $\rho$ is the thermal density matrix and $\bket{\rho^{1/2}}$ is the thermofield double state (TFD). However, they still define the ``size'' of $\ket{\Psi}$ relative to the infinite-temperature TFD. We will adjust their definition as follows. Consider the free-fermion state $\vr$ (referred to as the ``naive model'' in~\cite{KS17-soft}) reproducing the Wightman function of the SYK model at a given temperature. The corresponding state $\bket{\vr^{1/2}}$ of the double system may be regarded as a fermionic vacuum. Then the size of a perturbed TFD is defined as the average number of quasiparticles with some energy-dependent weights.

At low temperatures, essential properties of the SYK model are governed by the Schwarzian action~\cite{Kit.KITP.2,MS16-remarks,KS17-soft}, which describes a black hole in $1+1$ dimensions~\cite{MSY16}. The relevant part of black hole physics goes back to the work of Dray and 't~Hooft~\cite{DtH85}, who discovered that an infalling object generates a gravitational ``shock wave'' at the past horizon. A classical shock affects quantum correlation functions between the two sides of an eternal black hole~\cite{ShSt13}. In fact, a two-sided black hole perturbed by a shock is analogous to the state $\ket{\Psi}$ we have previously discussed. 'T~Hooft also proposed a fully quantum model of gravitational shocks, which mediate the scattering between incoming and outgoing particles~\cite{tH90}. The scattering amplitude is $a(p^{\mathrm{out}},p^{\mathrm{in}}) =\exp(ifp^{\mathrm{out}}p^{\mathrm{in}})$, where $p^{\mathrm{out}}$, $p^{\mathrm{in}}$ are the particles' null momenta and $f$ depends on their angular coordinates. Since the states of particles inside a black hole are inaccessible, the physical meaning of this theory is rather elusive and was understood long after 't~Hooft's work. If one wants to operate with quantum observables only on one side, the relevant quantity is an out-of-time-order correlator. Such correlators were calculated by Shenker and Stanford for black holes~\cite{ShSt14} and by Maldacena, Stanford, and Yang for the SYK model at low temperature~\cite{MSY16}. This includes the nonlinear, late-time regime. Stanford, Yang, and Yao~\cite{SYY21} conjectured that the general form of these solutions is valid for large-$N$ models at any temperature. We will show it is indeed the case.

A surprising connection exists between OTOCs and two-point functions. In a nutshell, chaotic dynamics contributes to equilibrium noise. Parker, Cao, Avdoshkin, Scaffidi, and Altman~\cite{PCASA18} conjectured a lower bound on high-frequency tails of the spectral function in terms of the Lyapunov exponent. They proved it at infinite temperature, and Avdoshkin and Dymarsky extended the derivation to finite temperatures~\cite{AvDy19}. We will give a different proof in section~\ref{sec:discussion}.

\section{The big picture}
\label{sec: big picture}
\begin{figure}\centering
\begin{tabular}{c@{\hspace{1.5cm}}c}
$\vcenter{\hbox{\includegraphics{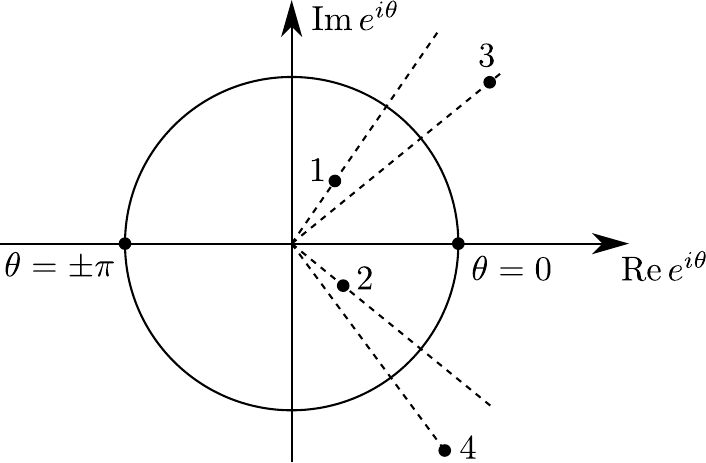}}}$ &
    \begin{tikzpicture}[baseline={(current bounding box.center)}]
    \draw[thick] (0pt,0pt) ellipse (13pt and 40pt);
    \filldraw[white, fill=white] (-13.2pt,6.55pt) rectangle (-10pt,9.45pt);
     \draw[thick] (-12.3pt, 10pt) -- (-72pt, 10pt);
    \draw[thick] (-12.5pt, 6pt) -- (-72pt,6pt);
    \draw[thick] (-72pt,10pt) arc (90:270:2pt);
    \filldraw (-74pt,8pt) circle (1.2pt) node[left] {$3$};
     \filldraw[white, fill=white] (-13.2pt,-6.55pt) rectangle (20pt,-9.45pt);
     \draw[thick] (12.9pt, -10pt) -- (-48pt, -10pt);
    \draw[thick] (13.2pt, -6pt) -- (-48pt,-6pt);
    \draw[thick] (-48pt,-10pt) arc (270:90:2pt);
    \filldraw (-50pt,-8pt) circle (1.2pt) node[left]{$4$};
      \filldraw[fill=white,white] (0.5pt,40pt) rectangle (6pt,37pt);
    \draw[thick](0pt,40pt) -- (60pt,40pt);
    \draw[thick](4.6pt,37pt) -- (60pt,37pt);
     \draw[thick] (60pt,40pt) arc (90:-90:1.5pt);
     \filldraw (61.5pt,38.5pt) circle (1.2pt) node[right]{$2$};
       \filldraw[fill=white,white] (0.5pt,-40pt) rectangle (6pt,-37pt);
    \draw[thick](0pt,-40pt) -- (60pt,-40pt);
    \draw[thick](4.6pt,-37pt) -- (60pt,-37pt);
    \draw[thick] (60pt,-40pt) arc (-90:90:1.5pt);
    \filldraw (61.5pt,-38.5pt) circle (1.2pt) node[right]{$1$};
    \end{tikzpicture}
\vspace{3pt}\\
(a) & (b)
\end{tabular}
\caption{(a) The complex times $\theta_1,\theta_2,\theta_3,\theta_4$ in terms of the variable $e^{i\theta}=e^{-t}e^{i\tau}$. (b) A symmetric case, where the points 1,2,3,4  are placed on a cylinder with the axial coordinate $t$ (going left to right) and connected by a contour such that the contour ordering coincides with the $\tau$ ordering.}
\label{fig:1234}
\end{figure}
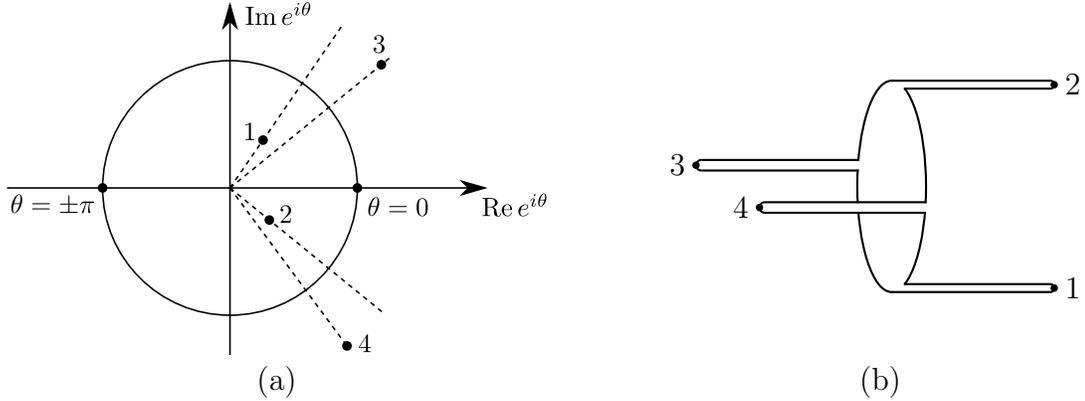

First, let us fix the notation and define the exact problem. We set $\beta=2\pi$ and consider complex times $\theta=\tau+it$ with $\tau$ defined modulo $2\pi$. In these units, the Lyapunov exponent satisfies the inequality $0<\kap<1$; the limits $\kap\to 0$ and $\kap\to 1$ can also be considered. To ensure the finiteness of correlation functions, operators should be ordered by $\tau$. The ordering involves fermionic parity, which will be indicated by a variable $\zeta=0,1$. Let $X_1,X_2$ have parity $\zeta_1=\zeta_2$ and let $X_3,X_4$ have parity $\zeta_3=\zeta_4$. The main function to study is this one:
\begin{equation}\label{OTOC0}
\begin{aligned}
\OTOC_{X_1,X_2,X_3,X_4}(\theta_1,\theta_2,\theta_3,\theta_4)
&=\bcorr{\TT X_1(\theta_1) X_2(\theta_2) X_3(\theta_3) X_4(\theta_4)}\\
&=(-1)^{\zeta_2\zeta_3}
\bcorr{X_1(\theta_1) X_3(\theta_3) X_2(\theta_2) X_4(\theta_4)},
\end{aligned}
\end{equation}
where (see also figure~\ref{fig:1234})
\begin{equation}\label{theta_t}
\theta_j=\tau_j+ i  t_j,\qquad
\tau_1 \geq \tau_3 \geq \tau_2 \geq \tau_4 \geq \tau_1-2\pi,\qquad
t_1\approx t_2 > t_3\approx t_4.
\end{equation}
Using the time translation symmetry, we can arrange that $t_1,t_2$ are positive, $t_3,t_4$ are negative, and all four times are approximately equal in magnitude. This allows for a symmetric interpretation of the OTOC, whereby $X_3(\theta_3)$ and $X_4(\theta_4)$ create a perturbation to the thermofield double that propagates forward in time, while $X_1(\theta_1)$ and $X_2(\theta_2)$ create a backward-propagating perturbation. These counter-propagating ``waves'' interact at times around $0$, where they are both relatively weak. More exactly, let us assume that
\begin{equation}
|t_1-t_2|,\,|t_3-t_4|\lesssim 1,\qquad
\frac{t_1+t_2}{2},\,-\frac{t_3+t_4}{2}\gg 1,\qquad
C^{-1}e^{(t_1+t_2)/2},\,C^{-1}e^{-(t_3+t_4)/2}\ll 1.
\end{equation}
The quantity relevant to the OTOC, $C^{-1}e^{(t_1+t_2-t_3-t_4)/2}$, is generally of the order of one.

We will show that the OTOC has the same general form as in the black hole problem~\cite{ShSt14}:
\begin{empheq}[box=\widebox]{align}
\label{OTOC1}
&\begin{aligned}
&\OTOC_{X_1,X_2,X_3,X_4}(\theta_1,\theta_2,\theta_3,\theta_4)\\
&\hspace{1.5cm}
=\int_{0}^{\infty}\!dy_\A \int_{0}^{\infty}\!dy_\R\,
e^{-\lambda y_{\A}y_{\R}}\,
h^\R_{X_1,X_2}\bigl(y_\A;\theta_1-\theta_2\bigr)\,
h^\A_{X_3,X_4}\bigl(y_\R;\theta_3-\theta_4\bigr),
\end{aligned}\\[5pt]
&\text{where}\qquad
\lambda= C^{-1}e^{i\kap(\pi-\theta_1-\theta_2+\theta_3+\theta_4)/2}.
\end{empheq}
Let us explain this result before deriving it. The integration variables $y_\A$ and $y_\R$ are analogous to the null momenta on the past and future horizons ($p^{\mathrm{out}}$, $p^{\mathrm{in}}$ in~\cite{tH90} or $p^u_1$, $p^v_2$ in~\cite{ShSt14}), which are due to particles moving and generating gravitational shocks along the opposite horizons. More exactly, $y_\A$ (representing a perturbation source in the future) and $y_\R$ (representing a similar source in the past) determine the magnitudes $z_\A$, $z_\R$ of the backward-propagating (advanced) and forward-propagating (retarded) modes, respectively:
\begin{equation}\label{z_vs_y}
z_\A=C^{-1}e^{-i\kap(\theta_1+\theta_2)/2}y_\A,\qquad
z_\R=C^{-1}e^{i\kap(\theta_3+\theta_4)/2}y_\R.
\end{equation}
The exponential factor in \eqref{OTOC1},
\begin{equation}\label{sc_ampl}
a(y_\A,y_\R) := e^{-\lambda y_{\A}y_{\R}} 
= \exp\bigl(-Ce^{i\kap\pi/2}z_{\A}z_{\R}\bigr),
\end{equation}
describes interaction between the counter-propagating modes and may be understood as a scattering amplitude.

For a sharper interpretation, let us assume that
\begin{equation}\label{spec_tau}
\tau_1=\tau_3=\frac{\tau}{2},\qquad
\tau_2=\tau_4=-\frac{\tau}{2},\qquad 0 \leq \tau \leq 2\pi
\end{equation}
so that $z_\A=C^{-1}e^{\kap(t_1+t_2)}y_\A$ and $z_\R=C^{-1}e^{-\kap(t_3+t_4)}y_\R$ are real and positive.\footnote{As a generalization, it is sufficient to assume that $\pi\geq\tau_1\geq\tau_3\geq 0 \geq\tau_2\geq\tau_4\geq-\pi$. In this case, one can arrange for $z_\A$, $z_\R$ given by Eq.~\eqref{z_vs_y} to be real and positive by taking the integrals over $y_\A,y_\R$ in \eqref{OTOC1} along certain paths in the complex plane.} In section~\ref{sec:size_operator}, we will interpret the quantities $Cz_\A$, $Cz_\R$ as eigenvalues of certain operators $Q_\A$, $Q_\R$ acting on the double system. Condition \eqref{spec_tau} also allows for the definition of in- and out-states and an $S$-matrix~\cite{KS17-soft,GuKi18}. However, let us take the idea of scattering informally and posit that the operators $X_2(\theta_2)$ and $X_1^\dag(\theta_1)$ create some local excitations (around the time $t_1\approx t_2$) dressed with a backward-propagating mode characterized by $y_\A$, and similarly, $X_4(\theta_4)$ and $X_3^\dag(\theta_3)$ create local excitations dressed with a forward-propagating mode characterized by $y_\R$. The complex numbers $h^\R_{X_1,X_2}(y_\A;\theta_1-\theta_2)$ and $h^\A_{X_3,X_4}(y_\R;\theta_3-\theta_4)$ represent the inner products between the local states projected onto the given values of $y_\A$ and $y_\R$. In particular, if $X_1^\dag=X_2$,\, $t_1=t_2$,\, $X_3^\dag=X_4$,\, $t_3=t_4$, then the functions
\begin{equation}\label{normalized_h}
w^\R(y_\A)=\frac{h^\R_{X_1,X_2}(y_\A;\tau)}{\corr{X_1(\tau)X_2(0)}},
\qquad
w^\A(y_\R)=\frac{h^\A_{X_3,X_4}(y_\R;\tau)}{\corr{X_3(\tau)X_4(0)}}
\end{equation}
are real, nonnegative, and properly normalized so that they may be interpreted as the probability distributions of $y_\A$, $y_\R$.

Perhaps the most important observation is concerned with the scattering amplitude \eqref{sc_ampl}. In the situation we have described, where $z_\A\geq 0$ and $z_\R\geq 0$, one finds that $|a(y_\A,y_\R)|=1$ if $\kap=1$. If $\kap<1$, the amplitude $a(y_\A,y_\R)$ has absolute value less than $1$. This is a signature of inelastic scattering due to the production of strings~\cite{ShSt14} or similar but simpler objects~\cite{GuKi18}.

\section{Derivation of the main equation}\label{sec:derivation}

We now derive equation \eqref{OTOC1}, postponing one step (namely, a matrix element interpretation of $h^\R$, $h^\A$, and related functions) until section~\ref{sec:melem}. First, let us consider the early-time OTOC. We assume that there is one fastest-growing scrambling mode (which comes in the retarded and advanced variants). This assumption amounts to the following ansatz~\cite{KS17-soft}:\footnote{Here, $C^{-1}$, $\VF^\R$, $\VF^\A$ are defined up to constant factors; only their product is fixed. By requiring positivity in certain cases (e.g.\ when $\theta_4=-\theta_3^*$ and $X_4=X_3^\dag$; see section~\ref{sec:VR}), the ambiguity is reduced down to positive factors. We also assume that $\VF^\R_{X_1,X_2}(\theta)$ and $\VF^\A_{X_3,X_4}(\theta)$ are of the order of $1$ if $\theta\sim 1$, whereas $C^{-1}$ is small.}
\begin{equation}\label{1-mode}
\begin{aligned}
\OTOC_{X_1,X_2,X_3,X_4}(\theta_1,\theta_2,\theta_3,\theta_4)
&\approx \corr{X_1(\theta_1)X_2(\theta_2)}\corr{X_3(\theta_3)X_4(\theta_4)}\\
&-\lambda\,
\VF^\R_{X_1,X_2}(\theta_1- \theta_2)\, \VF^\A_{X_3,X_4}(\theta_3-\theta_4),
\end{aligned}
\end{equation}
where $\lambda= C^{-1}e^{i\kap(\pi-\theta_1-\theta_2+\theta_3+\theta_4)/2}$. The approximation is valid if $t:=\frac{t_1+t_2-t_3-t_4}{2}$ is sufficiently large such that any decaying or more slowly growing modes may be neglected. We also suppose that $|\lambda|=C^{-1}e^{\kap t}$ is much less than $1$.

For the SYK model with Majorana operators $\chi_1,\dots,\chi_N$, equation \eqref{1-mode} is specialized as follows:
\begin{equation}\label{1-mode-SYK}
\frac{1}{N^2}\sum_{j,k}
\bcorr{\TT \chi_j(\theta_1)\chi_j(\theta_2)\chi_k(\theta_3)\chi_k(\theta_4)}
\approx \corr{\chi(\theta_1)\chi(\theta_2)}\corr{\chi(\theta_3)\chi(\theta_4)}
-\lambda\,\VF^\R(\theta_1- \theta_2)\,\VF^\A(\theta_3-\theta_4).
\end{equation}
The coefficient $C$ in this case is of the order of $N$. (At low temperatures, $C\sim\frac{N}{\beta J}$~\cite{MS16-remarks}, but we assume that $\beta$ and $J$ are fixed, whereas $N\to\infty$.) The second term in \eqref{1-mode-SYK} is given by a sum of ladder diagrams. Let us replace it with a schematic drawing:
\begin{equation}
\begin{tikzpicture}
\node[svertex] (R) at (-25pt,0pt) {};
\node[svertex] (A) at (25pt,0pt) {};
\draw[thick] (R) -- ++(135:12pt) node[left]{\scriptsize$1$};
\draw[thick] (R) -- ++(-135:12pt) node[left]{\scriptsize$2$};
\draw[thick] (A) -- ++(45:12pt) node[right]{\scriptsize$3$};
\draw[thick] (A) -- ++(-45:12pt) node[right]{\scriptsize$4$};
\draw[wavy] (A) to (R);
\end{tikzpicture}
\,=\lambda\,\VF^\R(\theta_1- \theta_2)\,\VF^\A(\theta_3-\theta_4),
\end{equation}
where the functions $\VF^\R$, $\VF^\A$ are associated with the two vertices and $\lambda$ (the scrambling mode propagator) with the wavy line. It is sometimes convenient to cut the wavy line in half:
\begin{equation}
\begin{tikzpicture}
\node[svertex] (R) at (-25pt,0pt) {};
\draw[thick] (R) -- ++(135:12pt) node[left]{\scriptsize$1$};
\draw[thick] (R) -- ++(-135:12pt) node[left]{\scriptsize$2$};
\draw[wavy] (R) to (0pt,0pt);
\end{tikzpicture}
\,=C^{-1}e^{-i\kap(\theta_1+\theta_2)/2}\,\VF^\R(\theta_1-\theta_2),
\qquad
\begin{tikzpicture}
\node[svertex] (A) at (25pt,0pt) {};
\draw[thick] (A) -- ++(45:12pt) node[right]{\scriptsize$3$};
\draw[thick] (A) -- ++(-45:12pt) node[right]{\scriptsize$4$};
\draw[wavy] (A) to (0pt,0pt);
\end{tikzpicture}
=C^{-1}e^{i\kap(\theta_3+\theta_4)/2}\,\VF^\A(\theta_3-\theta_4).
\end{equation}
To glue these pieces back together, one needs to insert the factor $Ce^{i\kap\pi/2}$. (The cutting and gluing of ladder diagrams was used in~\cite{GuKi18} to derive an expression for the coefficient $C$ in terms of the retarded kernel, the elementary unit of such ladders.)

\begin{figure}
\centering\includegraphics{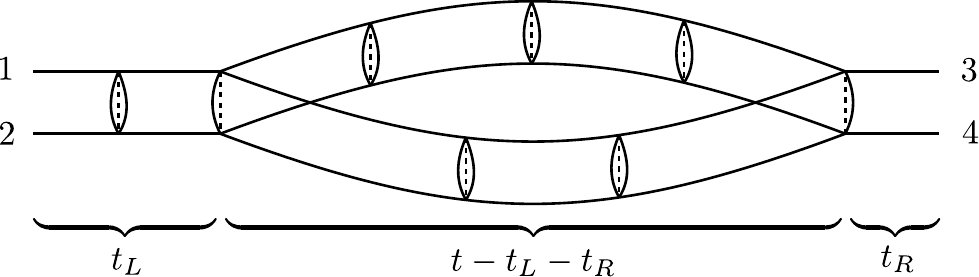}
\caption{An SYK diagram contributing to the OTOC at late times, such that $\lambda\sim 1$.}
\label{fig:double_ladder}
\end{figure}

When $\lambda\sim 1$, simple ladders are not sufficient, and we have to include more complex diagrams such as shown in figure~\ref{fig:double_ladder}. This particular diagram has an additional small factor $N^{-1}\sim C^{-1}$ due to branching, and the overall value can be estimated as follows:
\begin{equation}
C^{-2}e^{\kap t_L}\bigl(e^{\kap(t-t_L-t_R)}\bigr)^2e^{\kap t_R}
=|\lambda|^2e^{-\kap(t_L+t_R)}.
\end{equation}
Therefore, the diagram is significant if $\kap(t_L+t_R)\lesssim 1$. In general, significant diagrams consist of parallel ladders that join near the initial and final times, with no branching in the middle. They may be depicted like this:
\begin{equation}\label{diagrams_OTOC}
\begin{tikzpicture}
\node[bvertex] (R) at (-30pt,0pt) {};
\node[bvertex] (A) at (30pt,0pt) {};
\draw[thick] (R) -- ++(135:20pt) node[left]{\scriptsize$1$};
\draw[thick] (R) -- ++(-135:20pt) node[left]{\scriptsize$2$};
\draw[thick] (A) -- ++(45:20pt) node[right]{\scriptsize$3$};
\draw[thick] (A) -- ++(-45:20pt) node[right]{\scriptsize$4$};
\draw[wavy] (A) to[out=140,in=40] (R);
\draw[wavy] (A) to (R);
\draw[wavy] (A) to[out=-140,in=-40] (R);
\end{tikzpicture}\,.
\end{equation}
Thus, we arrive at a nonlinear generalization~\cite{GuKi18} of the single-mode ansatz:
\begin{equation}\label{eq:nlOTOC}
\OTOC_{X_1,X_2,X_3,X_4}(\theta_1,\theta_2,\theta_3,\theta_4)
= \sum_{m=0}^{\infty}\frac{(-\lambda)^m}{m!}\,
\VF_{X_1,X_2}^{\R,m}(\theta_1-\theta_2)\,
\VF_{X_3,X_4}^{\A,m}(\theta_3-\theta_4).
\end{equation}
In particular, $\VF_{X_1,X_2}^{\R,0}(\theta_1-\theta_2) =\corr{X_1(\theta_1)X_2(\theta_2)}$ and $\VF_{X_1,X_2}^{\R,1} =\VF_{X_1,X_2}^{\R}$. The expansion~\eqref{eq:nlOTOC} is only asymptotic; see section~\ref{sec:large-q_OTOC} for an illustration.\medskip

The next step is to cut all wavy lines in half, perform an independent summation on each side, and figure how to combine the results. For example, on the left side, we get:
\begin{equation}
F^\R_{X_1,X_2}(z;\theta_1,\theta_2) =\,
\begin{tikzpicture}
\draw[thick] (0pt,-10pt) node[left]{\scriptsize$2$}
to[out=45,in=-45] (0pt,10pt) node[left]{\scriptsize$1$};
\end{tikzpicture}\;
+\,
\begin{tikzpicture}
\node[svertex] (R) at (-25pt,0pt) {};
\node[sdot] (D) at (0pt,0pt) {};
\draw[thick] (R) -- ++(135:16pt) node[left]{\scriptsize$1$};
\draw[thick] (R) -- ++(-135:16pt) node[left]{\scriptsize$2$};
\draw[wavy] (D) to (R);
\end{tikzpicture}\;
+\,
\begin{tikzpicture}
\node[mvertex] (R) at (-30pt,0pt) {};
\node[sdot] (D1) at (0pt,10pt) {};
\node[sdot] (D2) at (0pt,-10pt) {};
\draw[thick] (R) -- ++(135:18pt) node[left]{\scriptsize$1$};
\draw[thick] (R) -- ++(-135:18pt) node[left]{\scriptsize$2$};
\draw[wavy] (D1) to[out=180,in=30] (R);
\draw[wavy] (D2) to[out=180,in=-30] (R);
\end{tikzpicture}\;
+\,
\begin{tikzpicture}
\node[bvertex] (R) at (-35pt,0pt) {};
\node[sdot] (D1) at (0pt,12pt) {};
\node[sdot] (D2) at (0pt,0pt) {};
\node[sdot] (D3) at (0pt,-12pt) {};
\draw[thick] (R) -- ++(135:20pt) node[left]{\scriptsize$1$};
\draw[thick] (R) -- ++(-135:20pt) node[left]{\scriptsize$2$};
\draw[wavy] (D1) to[out=180,in=40] (R);
\draw[wavy] (D2) to (R);
\draw[wavy] (D3) to[out=-180,in=-40] (R);
\end{tikzpicture}\;
+\;\cdots
\end{equation}
More exactly, the two one-sided sums are defined as follows:
\begin{equation}\label{FRA}
\begin{aligned}
F_{X_1,X_2}^\R(z;\theta_1,\theta_2) &=\sum_{m=0}^{\infty}
\frac{\bigl(-e^{-i\kap(\theta_1+\theta_2)/2}z\bigr)^m}{m!}\,
\VF_{X_1,X_2}^{\R,m}(\theta_1-\theta_2),\\[3pt]
F_{X_3,X_4}^\A(z;\theta_3,\theta_4) &=\sum_{m=0}^{\infty} 
\frac{\bigl(-e^{i\kap(\theta_3+\theta_4)/2}z\bigr)^m}{m!}\,
\VF_{X_3,X_4}^{\A,m}(\theta_3-\theta_4).
\end{aligned}
\end{equation}
Here, $z$ is an abstract parameter that represents the strength of some mean-field perturbation to the thermofield double. (In the big picture we are trying to justify, the argument of $F^\R$ is $e^{i\kap\pi/2}z_\R$ and the argument of $F^\A$ is $e^{i\kap\pi/2}z_\A$.) Thus, $F_{X_1,X_2}^\R$, $F_{X_3,X_4}^\A$ are correlation functions on the perturbed background; they can be found by solving mean-field equations on the double Keldysh contour. The linear version of this problem has been studied extensively, see e.g.~\cite{Kit.KITP.1,MS16-remarks,MSW17,GuKi18}. The nonlinear equations have been solved for various weakly coupled models where they can be reduced to an analogue of the Boltzmann equation~\cite{AFI16}. For the SYK model, these equations have been written explicitly (but not solved) in~\cite{ZGK20}. We will obtain an analytic solution in the large-$q$ limit and numerical solutions for $q=4,6$.

As will be seen in concrete examples, the Taylor series \eqref{FRA} have finite convergence radius. However, the path to our goal, equation~\eqref{OTOC1}, lies through inverse Laplace transform, which requires analytic continuation. To this end, we will show in section~\ref{sec:melem} that the function values $F_{X_1,X_2}^\R(z;\theta_1,\theta_2)$ and $F_{X_3,X_4}^\A(z;\theta_3,\theta_4)$ can be represented as matrix elements of $\exp(-zQ_{\A})$ and $\exp(-zQ_{\R})$, respectively, where $Q_\A$ and $Q_\R$ are some positive-semidefinite operators acting on the double system.\footnote{Quantum states of the double system, such as $\ket{A}$, correspond to operators acting on the single system (in this case, $A$). Furthermore, $\braket{A}{B}=\Tr(A^{\dag}B)$; see section~\ref{sec:TFD} for exact rules.} For example,
\begin{equation}\label{FR_melem}
F^\R_{X_1,X_2}(z;\theta_1,\theta_2)
= \bbra{X_1(\theta_1)^\dag\rho^{1/2}}
e^{-zQ_\A}
\bket{X_2(\theta_2)\rho^{1/2}}\quad\;
\text{for }\, \pi\geq\Re\theta_1\geq 0\geq\Re\theta_2\geq -\pi.
\end{equation}
The conditions on $\theta_1,\theta_2$ guarantee that the states $\ket{X_1(\theta_1)^\dag\rho^{1/2}}$ and $\ket{X_2(\theta_2)\rho^{1/2}}$ have bounded norm even in the $N\to\infty$ limit. (In section~\ref{sec:discussion}, these conditions will be relaxed, leading to a bound on the high-frequency decay of the spectral function.) Furthermore, $\|e^{-zQ_\A}\|\leq 1$ for $\Re z\geq 0$ because $Q_\A$ is positive-semidefinite. It follows that for given $\theta_1$ and $\theta_2$, the matrix element \eqref{FR_melem} is analytic in the half-plane $\Re z> 0$, tending to $0$ as $\Re z\to+\infty$. The function $F_{X_3,X_4}^\A(z;\theta_3,\theta_4)$ has similar analytic properties.

For convenience, let us absorb the common factors in \eqref{FRA} into the variable $z$:
\begin{equation}\label{F_vs_f}
\begin{alignedat}{2}
F_{X_1,X_2}^\R(z;\theta_1,\theta_2)
&=f_{X_1,X_2}^\R\bigl(e^{-i\kap(\theta_1+\theta_2)/2}z;\,
\theta_1-\theta_2\bigr),\quad\: &
f_{X_1,X_2}^\R(x;\theta)
&=\sum_{m=0}^{\infty}
\frac{(-x)^m}{m!}\,\VF_{X_1,X_2}^{\R,m}(\theta),\\[3pt]
F_{X_3,X_4}^\A(z;\theta_3,\theta_4)
&=f_{X_3,X_4}^\A\bigl(e^{i\kap(\theta_3+\theta_4)/2}z;\,
\theta_3-\theta_4\bigr),\quad\: &
f_{X_3,X_4}^\A(x;\theta)
&=\sum_{m=0}^{\infty} 
\frac{(-x)^m}{m!}\,\VF_{X_3,X_4}^{\A,m}(\theta).
\end{alignedat}
\end{equation}
The new functions $f_{X_1,X_2}^\R(x;\theta)$ and $f_{X_3,X_4}^\A(x;\theta)$ are analytic in $x$ for $|\arg x|< \frac{\pi}{2}+\phi$, where
\begin{equation}\label{eq:analytical_pre}
\phi=\frac{\kap}{2}\min\{\tau,2\pi-\tau\},\quad \tau=\Re\theta,
\end{equation}
and tend to $0$ as $|x|$ goes to infinity while $\arg x$ remaining constant. Thus, they can be expressed using the Laplace transform,
\begin{equation}\label{Laplace}
f_{X_1,X_2}^\R(x;\theta)=\int_0^{\infty}e^{-xy}\,h_{X_1,X_2}^\R(y;\theta)\,dy,
\qquad
f_{X_3,X_4}^\A(x;\theta)=\int_0^{\infty}e^{-xy}\,h_{X_3,X_4}^\A(y;\theta)\,dy,
\end{equation}
where $h_{X_1,X_2}^\R(y;\theta)$ and $h_{X_1,X_2}^\A(y;\theta)$ are analytic in $y$ in the domain $|\arg y|<\phi$.

Now, using the definition of $f_{X_1,X_2}^\R$, $f_{X_3,X_4}^\A$ and equation \eqref{Laplace}, we find that
\begin{equation}\label{moments}
\VF_{X_1,X_2}^{\R,m}(\theta)=\int_0^{\infty}y^{m}\,h_{X_1,X_2}^\R(y;\theta)\,dy,
\qquad
\VF_{X_3,X_4}^{\A,m}(\theta)=\int_0^{\infty}y^{m}\,h_{X_3,X_4}^\A(y;\theta)\,dy.
\end{equation}
Plugging these expressions into \eqref{eq:nlOTOC}, we obtain equation \eqref{OTOC1}.

\section{Small perturbations to the thermofield double}\label{sec:linear}

In the linear (early-time) regime, scrambling modes are treated as small perturbations to the thermofield double state. A mode is generally in a superposition of mean-field states of different amplitudes, but in the linear approximation, only the average amplitude $c$ and the corresponding change in the two-point function $\delta\corr{\chi_\alpha(\tau_1)\chi_{\beta}(\tau_2)}\propto c$ matter. (Here, $\alpha,\beta,\ldots$ refer to Majorana modes in an SYK-like setting.)

We find it convenient to use a Hilbert space formulation of mean-field theory, known as the ``naive model''~\cite{KS17-soft}. Formally, it is just a representation of the two-point function by a Gaussian fermionic state. To give it more meaning, consider the operators $\chi_\alpha(it)$ with $t$ in an interval much shorter than the scrambling time. Their thermal expectation values obey Wick's theorem with $1/N$ accuracy, and so it is tempting to say that the quantum state is almost Gaussian. This is actually not correct because $1/N$ errors add up; for example, one cannot calculate the expectation value of the SYK Hamiltonian $\sum_{j<k<l<m}J_{jklm}\chi_j\chi_k\chi_l\chi_m$ using Wick's theorem. (The accumulation of errors can be illustrated by constructing $N$ linear combinations of basic Majorana operators that almost anticommute pairwise but not in the bigger set. Indeed, there exist $N$ unit vectors with $O(1/N)$ mutual inner products that are nonetheless linearly dependent.) To avoid this problem, we assume that $\alpha,\beta,\ldots$ belong to a small subset of modes. Then the operators $\chi_\alpha(it)$ may be regarded as free-fermion, i.e. $\{\chi_\alpha(it),\chi_\beta(it')\}$ is just a number with acceptable accuracy. Furthermore, the restriction of the thermal state to the subalgebra generated by these operators is Gaussian. For the double system and similarly doubled subalgebra, the TFD is reduced to a pure Gaussian state, which may be called a free-fermion vacuum.

In this section, we mathematically define the double system and consider the perturbed TFD. Although it is still a pure state, its restriction $R$ to the aforementioned subalgebra is not. We examine the structure of $R$ and decompose it into a coherent and an incoherent part. (This is the only task that requires the use of the naive model, and it is tangential to our main goal.) Then we introduce ``size operators'' $Q_\R$, $Q_\A$ that measure the magnitudes of forward-growing (retarded) and backward-growing (advanced) scrambling modes, respectively. These operators are positive-semidefinite, vanish on the TFD, and are sensitive only to the incoherent perturbation parts, which are related to the commutator OTOC.

\subsection{The double system}\label{sec:TFD}

Let $\calH$ be the Hilbert space of a physical system, and let us introduce its almost identical copy with the Hilbert space $\calH^*$, that is, the space of linear functional on $\calH$ (or equivalently, bra-vectors). The algebra $\LL(\calH)$ of operators acting on $\calH$ is canonically isomorphic to $\calH\otimes\calH^*$. Thus, any operator $A\in\LL(\calH)$ may be interpreted as a vector $\ket{A}\in\calH\otimes\calH^*$. In particular, $|{\rm TFD} \rangle= | \rho^\frac{1}{2} \rangle$ corresponds to the square root of the thermal density matrix $\rho = Z^{-1} e^{-\beta H}$. The Hermitian inner product on $\calH\otimes\calH^*$ has the following expression in the operator language:
\begin{equation}
\braket{A}{B}=\Tr(A^{\dag}B).
\end{equation}

Now, we consider operators acting on the double system. In the bosonic case, the formalism is quite simple. Note that the operator algebra $\LL(\calH^*)$ is isomorphic to $\LL(\calH)$ with the multiplication order reversed; this new algebra is denoted by $\LL(\calH)^{\rm op}$. Thus, $\LL(\calH\otimes \calH^*)\cong \LL(\calH) \otimes\LL(\calH)^{\rm op}$. An element $X\bdot Y\in \LL(\calH) \otimes \LL(\calH)^{\rm op}$ acts as follows:
\begin{equation}
(X\bdot Y)\ket{A}= \ket{XAY}.
\end{equation} 
In the $\ZZ_2$-graded case, we define the left and right actions of $\LL(\calH)$ on itself so as to produce a representation of the $\ZZ_2$-graded algebra product $\LL(\calH) \otimes_{\rm gr} \LL(\calH)^{\rm op}$:
\begin{equation}
\begin{aligned}
&(X\bdot I) \ket{A} = \ket{XA} \qquad \text{(left action)}, \\
&(I\bdot Y) \ket{A} = i^{\deg(A Y)-\deg(A) } \ket{AY} \qquad \text{(right action)}.
\end{aligned}
\label{LR action}
\end{equation} 
Here, the phase factor $ i^{\deg(A Y)-\deg(A) }$ is designed to ensure that the left and right actions commute up to $(-1)^{\deg X \cdot \deg Y}$. For a general element  $X\bdot Y = (X \bdot I) (I \bdot Y)$ of $\LL(\calH) \otimes_\gr \LL(\calH)^{\rm op}$, we have 
\begin{equation}
\begin{aligned}
(X_1\bdot Y_1) (X_2 \bdot Y_2)
&= (-1)^{\deg Y_1 \cdot \deg X_2} (X_1 X_2 \bdot Y_2 Y_1), \\
(X \bdot Y)^\dagger
&= (-1)^{\deg X \cdot \deg Y} (X^\dagger \bdot Y^\dagger).
\end{aligned}
\end{equation}

For an arbitrary $X\in \LL(\calH)$, we define an ``annihilation operator'' $a(X)$ such that $a(X)\ket{\rho^{\frac{1}{2}}} =0$:
\begin{equation}\label{Defa}
a(X) = \rho^{\frac{1}{4}} X \rho^{-\frac{1}{4}} \bdot I - i^{-\deg X} I \bdot \rho^{-\frac{1}{4}} X \rho^{\frac{1}{4}},
\end{equation}
where the phase factor $i^{-\deg X}$ is consistent with the choice in \eqref{LR action}. Note that
\begin{equation}\label{a_vs_grcomm}
a(X)\,\ket{\rho^{\frac{1}{4}}Y\rho^{\frac{1}{4}}}
=\bket{\rho^{\frac{1}{4}}[X,Y]_\gr\rho^{\frac{1}{4}}},
\end{equation}
where
\begin{equation}
[X,Y]_\gr = XY - (-1)^{\deg X\cdot\deg Y}YX.
\medskip
\end{equation}

Now, let $\chi_\alpha(it)$ be some Majorana fields satisfying Wick's theorem. (The example of interest is $\chi(it)=\chi_j(it)$ for a fixed site $j$ of the SYK model, considered at the time scale of the order of $1$ so that $1/N$ corrections are not exponentially amplified.) We work in the operator algebra generated by $\chi_\alpha(it)$, or equivalently, by $\tchi_\alpha (\omega) =\int_{-\infty}^{+\infty} \chi_\alpha (it) e^{i\omega t} dt$ with the commutation relations $\{\tchi_\alpha (\omega), \tchi_\beta (\omega')\} =A_{\alpha\beta}(\omega)\,2\pi \delta(\omega+\omega')$, where the spectral function $A$ also determines the correlation function:
\begin{equation}
\langle \tchi_\alpha (\omega) \tchi_\beta (\omega') \rangle
= \frac{A_{\alpha\beta}(\omega)}{1+e^{-\beta \omega}}\,
2\pi \delta (\omega+\omega').
\label{chiA}
\end{equation}
The thermal state $\rho$ restricted to this subalgebra will be denoted by $\vr$. In most cases, $\vr$ and $\rho$ can be used interchangeably; for example,
\begin{equation}
\vr^{-s} \tchi_{\alpha }(\omega )\vr^s
=\rho^{-s} \tchi_{\alpha }(\omega )\rho^s
= e^{s\beta H} \tchi_\alpha (\omega) e^{-s\beta H}
= e^{-s \beta \omega }\tchi_\alpha (\omega).
\end{equation}
The distinction between $\vr$ and $\rho$ will be important in the next section, when we will consider perturbations.

The next piece of formalism is useful to describe the structure of the perturbed TFD. For this purpose, it is convenient to represent $\chi_{\alpha }(it)$ as a delayed signal coming from the ``past horizon'', i.e.\ a standard heat bath with a flat spectral function. More exactly (see appendix~\ref{sec: horizon rep} for detail),
\begin{equation}\label{pasthor}
\tchi_\alpha (\omega)
= \sum_\beta L_{\alpha \beta}^\R(\omega) \tpsi_\beta (\omega),
\end{equation}
where $\{\tpsi_\alpha(\omega),\tpsi_{\beta}(\omega')\} =\delta_{\alpha\beta}\,2\pi \delta(\omega+\omega')$. The last condition is equivalent to the equation $L^\R(\omega) L^\R (\omega)^\dagger = A(\omega)$, and we also require that $L^\R(\omega)$ admits an analytic continuation to the upper half-plane and has reasonable asymptotic behavior at $\omega\to\infty$. Due to the analyticity condition, $\tpsi_\alpha(\omega)$ is defined for complex $\omega$ with positive imaginary part by inverting the transformation \eqref{pasthor} and using the regularization $\tchi_\alpha(\omega)\approx\int_{-\infty}^{+\infty} \chi_\alpha (it) e^{i\omega t} u(t)\,dt$, where $u$ is a broad Gaussian. But if $\Im\omega<0$, then $\tpsi_\alpha(\omega)$ is hard to express in terms of the original operators of the model. Nonetheless, we will use it as a formal construct in equations~\eqref{coh_part}, \eqref{incoh_part} for quantum states. This is legitimate because by definition, a state is a linear functional on observables. If one uses this definition directly, as in~\eqref{coh_part0}, \eqref{incoh_part0}, the past horizon representation is not necessary, though it helps avoid carrying the spectral function around. In the derivation of our main result, we will not use it at all.

The canonical annihilation operators for the double system are defined as follows:
\begin{gather}
\label{naivea}
\ta_\alpha (\omega)
= \frac{\Gamma \left( \frac{1}{2} - i \frac{\beta\omega}{2\pi} \right)}
{\sqrt{2\pi}}
\Bigl( {\underbrace{\vr^\frac{1}{4} \tpsi_\alpha (\omega) \vr^{-\frac{1}{4}}}
_{e^{\beta\omega/4}\tpsi_\alpha(\omega)}} \bdot I
+ i I \bdot
{\underbrace{\vr^{-\frac{1}{4}} \tpsi_\alpha (\omega) \vr^{\frac{1}{4}}}
_{e^{-\beta\omega/4}\tpsi_\alpha(\omega)}} \Bigr),
\displaybreak[0]\\[3pt]
\ta_\alpha (\omega) | \vr^{\frac{1}{2}} \rangle =0 , \quad \ta_{\alpha}(\omega^*)^\dagger| \vr^{\frac{1}{2}} \rangle = \frac{\sqrt{2\pi}}{\Gamma \left(\frac{1}{2}- i \frac{\beta\omega}{2\pi}\right)}\, \bket{\vr^{\frac{1}{4}} \tpsi_\alpha(-\omega) \vr^{\frac{1}{4}}},
\end{gather}
where the overall factor is chosen such that $\{\ta_\alpha(\omega), \ta_\beta(\omega')^\dagger\} = \delta_{\alpha\beta}\, 2\pi \delta(\omega-\omega')$ when $\omega$, $\omega'$ are real. In the OTOC analysis, we will use a variant of this equation:
\begin{equation}
\bigl\{ \ta_\alpha \bigl(\omega+ i \tfrac{\varkappa }{2}\bigr),\,   \ta_\beta \bigl(\omega'- i \tfrac{\varkappa}{2}\bigr)^\dagger \bigr\} = \delta_{\alpha\beta}\, 2\pi \delta (\omega-\omega')\qquad
\text{for }\, \omega,\omega'\in \RR.
\end{equation}

\subsection{Retarded vertex function as a deformation of the TFD}\label{sec:VR}

Let us again set $\beta=2\pi$ and consider fermionic operators $X_j$ at complex times $\theta_j=\tau_j+it_j$ with
\begin{equation}\label{eqn: equal time}
\tau_1= \pi ,\quad \tau_3=\frac{\pi}{2}, \quad
\tau_2=0,\quad \tau_4=-\frac{\pi}{2},\qquad
t_1\approx t_2 > t_3\approx t_4.
\end{equation}
This configuration is shown in Fig.~\ref{fig:1234}b. It contains a pair of operators in the future (namely, $X_1$ and $X_2$) and another pair in the past ($X_3$ and $X_4$). We may interpret either pair as a source that perturbs the thermofield double $\ket{\rho^{1/2}}$, whereas the other pair measures the resulting effect. Here, we choose to regard $X_3$ and $X_4$ as sources and interpret the OTOC as a matrix element of a combination of $X_1$, $X_2$ acting on the double system:
\begin{equation}\label{OTOC_melem}
\OTOC_{X_1,X_2,X_3,X_4}(\theta_1,\theta_2,\theta_3,\theta_4)
=\bbra{\rho^{1/4}X_4(it_4)^\dag\rho^{1/4}}
-iX_1(it_1)\bdot X_2(it_2)
\bket{\rho^{1/4}X_3(it_3)\rho^{1/4}}.
\end{equation}
We assume that $t_1\approx t_2\approx 0$, whereas $t_3,t_4$ are large and negative. Thus, we focus on the perturbation effect while pushing the sources to distant past (or distant future in other cases). For simplicity, let $X_4=X_3^\dag$ and $t_4=t_3$; then the above expression becomes (up to an overall factor) the expectation value of $-iX_1(it_1)\bdot X_2(it_2)$ on the pure state $\ket{\Psi}\bra{\Psi}$, where
\begin{equation}
\ket{\Psi}=b^{-1/2}\bket{\rho^{1/4}X_3(it_3)\rho^{1/4}},\qquad
b=\bbraket{\rho^{1/4}X_3(it_3)\rho^{1/4}}{\rho^{1/4}X_3(it_3)\rho^{1/4}}
=\corr{X_3(\theta_3)X_4(\theta_4)}.
\end{equation}
If $X_1=\chi_\alpha$ and $X_2=\chi_\beta$ are elementary Majorana operators, we may restrict $\ket{\Psi}\bra{\Psi}$ to the previously defined subalgebra, i.e.\ trace out the other degrees of freedom. (This amounts to using the ``naive model'' when detecting the TFD deformation.) Thus, we obtain a mixed state $R$ of a system of free fermions, and the OTOC is expressed as follows:
\begin{equation}
b^{-1}\OTOC_{\chi_\alpha,\chi_\beta,\cdots,\cdots}(\pi+it_1,it_2,\cdots,\cdots)
=\Tr\bigl((-i\chi_\alpha(it_1)\bdot \chi_\beta(it_2))\,R\bigr).
\end{equation}
The expression on the right-hand side is a perturbed version of the Wightman function,
\begin{equation}
\corr{\chi_\alpha(\pi+it_1)\chi_\beta(it_2)}
=\Tr\bigl((-i\chi_\alpha(it_1)\cdot \chi_\beta(it_2))\,\ket{\vr^{1/2}}\bra{\vr^{1/2}}\bigr).
\end{equation}

In this section, we study the deformation of the TFD in the linear order. Thus, the OTOC is given by equation \eqref{1-mode}, which can be simplified as follows:
\begin{equation}\label{pert}
b^{-1}\OTOC_{\chi_\alpha,\chi_\beta,\cdots,\cdots}(\pi+it_1,it_2,\cdots,\cdots)
=\corr{\chi_\alpha(\pi+it_1)\chi_\beta(it_2)}
- c\,e^{\varkappa \frac{t_1+t_2}{2}} \VF^\R_{\alpha\beta}(\pi+i(t_1-t_2)).
\end{equation}
The coefficient $c$ depends on the deformation source, i.e.\ on $X_3$ and $t_3$, but we may assume that $c>0$ by fixing the normalization of $\VF^\R$. The Fourier transform of this function, $\tVF^\R(\omega) =\int_{-\infty}^{+\infty}\VF^\R(\pi+it)e^{i\omega t}dt$, can be shown to satisfy the following relations:
\begin{equation}\label{symProp}
\tVF^\R(\omega) = \tVF^\R(-\omega)^T
= \tVF^\R(-\omega)^* = \tVF^\R(\omega)^\dagger,\qquad
\tVF^\R(\omega)\geq 0.
\end{equation}
The last inequality can be derived from the positivity of the incoherent deformation part, defined below, which is proportional to $\cos\frac{\pi\kap}{2}$ and related to commutator OTOCs.

The last term in \eqref{pert} characterizes the deformation of the free-fermion vacuum:
\begin{equation}\label{eq:pertcorr}
\delta\corr{-i\chi_\alpha(it_1)\bdot \chi_\beta(t_2)}
:= \Tr\bigl( (-i\chi_\alpha(t_1)\bdot\chi_\beta(t_2))\,\delta R\bigr)
= - c\,e^{\varkappa \frac{t_1+t_2}{2}} \VF^\R_{\alpha\beta}(\pi+i(t_1-t_2)),
\end{equation}
where
\begin{equation}
\delta R= R - \ket{\vr^{1/2}}\bra{\vr^{1/2}}.
\end{equation} 
An arbitrary infinitesimal deformation of a pure state can be decomposed into a ``coherent'' and an ``incoherent'' part:
\begin{equation}
\delta R = \ket{\xi}\bra{\vr^{1/2}} + \ket{\vr^{1/2}}\bra{\xi}
+ \sum_j \bigl(\ket{\eta_j}\bra{\eta_j}
- \braket{\eta_j}{\eta_j} \ket{\vr^{1/2}}\bra{\vr^{1/2}}\bigr),\qquad
\ket{\xi},\ket{\eta_j}\perp\ket{\vr^{1/2}}.
\end{equation}
In the case of a free-fermion system, $|\xi\rangle$ is a linear combination of $a^\dag_\alpha a^\dag_\beta |\vr^{1/2}\rangle$, whereas $|\eta_j\rangle$ is a combination of $a^\dag_\alpha|\vr^{1/2}\rangle$. To compute $|\xi\rangle$ and $|\eta_j\rangle$, we will consider $\delta\langle \ta_\alpha (\omega+i\frac{\varkappa}{2})\, \ta_\beta (\omega'+i\frac{\varkappa}{2}) \rangle$ and $\delta \langle \ta_\alpha(\omega+i\frac{\varkappa}{2})^\dag \ta_\beta (\omega'+i\frac{\varkappa}{2}) \rangle$, respectively.

Since the annihilation and creation operators are defined in terms of the ``past horizon'' operators $\tpsi_\beta(\omega)$ (see Eq.~\eqref{pasthor}), the first calculation step is to pass from $\delta\corr{-i\chi_\alpha(\pi+it_1) \bdot \chi_\beta(t_2)}$ in \eqref{eq:pertcorr} to
\begin{equation}
\delta\bcorr{\tpsi_\alpha\bigl(\omega+i\tfrac{\varkappa}{2}\bigr)\bdot
\tpsi_\beta\bigl(\omega'+i\tfrac{\varkappa}{2}\bigr)} =-ic\,\tXi^\R_{\alpha\beta}(\omega)\,\delta(\omega+\omega'),
\end{equation}
where
\begin{equation}
\tXi^\R(\omega) = L^\R\bigl(\omega+i\tfrac{\varkappa}{2}\bigr)^{-1}\,
\tVF^\R(\omega)
\left(L^\R\bigl(\omega+i\tfrac{\varkappa}{2}\bigr)^{-1}\right)^\dagger.
\end{equation}
The function $\tXi^\R(\omega)$ has properties similar to \eqref{symProp}.

The next step is to express the products of $a$ and $a^\dag$ using the definition \eqref{naivea}. Each such expression has four terms, but it is sufficient to consider only the terms with $\tpsi_\alpha$ and $\tpsi_\beta$ on different sides. (Indeed, the expectation value of any one-sided observable is not affected by the deformation.) Thus,
\begin{equation}
\begin{aligned}\label{coh_part0}
\hspace{20pt}&\hspace{-20pt}
\delta \bcorr{\ta_\alpha \bigl(\omega+i\tfrac{\kap}{2}\bigr)
\ta_\beta\bigl(\omega'+i\tfrac{\kap}{2}\bigr)}\\[2pt]
&= \frac{\Gamma(\frac{1}{2}+\frac{\kap}{2} - i \omega)  \Gamma(\frac{1}{2}+\frac{\kap}{2} - i \omega')   }{2\pi}
\begin{aligned}[t]
\Bigl(\; &ie^{\pi(\omega-\omega')/2}\,
\delta \bcorr{\tpsi_\alpha\bigl(\omega+i\tfrac{\kap}{2}\bigr)
\bdot \tpsi_\beta\bigl(\omega'+i\tfrac{\kap}{2})}\\
{}-{} &ie^{\pi(\omega'-\omega)/2}\,
\delta \bcorr{\tpsi_\beta \bigl(\omega'+i\tfrac{\kap}{2}\bigr)
\bdot \tpsi_\alpha \bigl(\omega+i\tfrac{\kap}{2}\bigr)}\Bigr)
\end{aligned}\\[2pt]
&= c\,
\bigl| \Gamma\bigl(\tfrac{1}{2}+\tfrac{\kap}{2}-i\omega\bigr) \bigr|^2\,
\delta(\omega+\omega')\,
2 \sinh(\pi\omega)\, \tXi^{\R}_{\alpha\beta} (\omega),
\end{aligned}
\end{equation} 
and similarly, 
\begin{equation}\label{incoh_part0}
\begin{aligned}
\hspace{20pt}&\hspace{-20pt}
\delta \bcorr{\ta_\alpha \bigl(\omega+i\tfrac{\kap}{2}\bigr)^\dag
\ta_\beta\bigl(\omega'+i\tfrac{\kap}{2}\bigr)}\\[2pt]
&= \frac{\Gamma(\frac{1}{2}+\frac{\kap}{2} - i \omega)  \Gamma(\frac{1}{2}+\frac{\kap}{2} - i \omega')   }{2\pi}
\begin{aligned}[t]
\Bigl(\; &ie^{\pi(\omega-\omega'-i\kap)/2}\,
\delta \bcorr{\tpsi_\alpha\bigl(-\omega+i\tfrac{\kap}{2}\bigr)
\bdot \tpsi_\beta\bigl(\omega'+i\tfrac{\kap}{2})}\\
{}+{} &ie^{\pi(\omega'-\omega+i\kap)/2}\,
\delta \bcorr{\tpsi_\beta \bigl(\omega'+i\tfrac{\kap}{2}\bigr)
\bdot \tpsi_\alpha \bigl(-\omega+i\tfrac{\kap}{2}\bigr)}\Bigr)
\end{aligned}\\[2pt]
&= c\,
\bigl| \Gamma\bigl(\tfrac{1}{2}+\tfrac{\kap}{2}-i\omega\bigr) \bigr|^2\,
\delta(\omega-\omega')\, 
2 \cos\bigl(\tfrac{\pi\kap}{2}\bigr)\, \tXi^{\R}_{\alpha\beta} (-\omega).
\end{aligned}
\end{equation} 
From this we obtain the expressions for the coherent part,
\begin{equation}\label{coh_part}
\begin{aligned}
|\xi \rangle &= c \int \frac{d\omega}{2\pi}\,
\frac{\bigl|\Gamma(\frac{1}{2} +\frac{\kap}{2} -i\omega)\bigr|^2}{2\pi}
\sinh(\pi\omega)
\sum_{\alpha,\beta} \tXi^{\R}_{\alpha \beta}(\omega)\,
\ta_\beta\bigl(-\omega-i\tfrac{\kap}{2}\bigr)^\dag
\ta_\alpha\bigl(\omega-i\tfrac{\kap}{2})^\dag\, \ket{\vr^{1/2}}\\
&= c \int \frac{d\omega}{2\pi}\, \sinh(\pi\omega)
\sum_{\alpha,\beta} \tXi^{\R}_{\alpha\beta}(\omega)\,
\bket{\vr^{1/4}\tpsi_\beta\bigl(\omega-i\tfrac{\kap}{2}\bigr)
\tpsi_\alpha\bigl(-\omega-i\tfrac{\kap}{2}\bigr)\vr^{1/4}},
\end{aligned}
\end{equation}
and for the incoherent part,
\begin{equation}\label{incoh_part}
\begin{aligned}
\sum_j | \eta_j \rangle \langle \eta_j |
&= c \cdot 2\cos\bigl(\tfrac{\pi\kap}{2}\bigr)
\int \frac{d\omega}{2\pi}\,     
\frac{\bigl|\Gamma(\frac{1}{2} + \frac{\kap}{2} - i\omega)\bigr|^2}{2\pi}
\sum_{\alpha,\beta}
\tXi^{\R}_{\alpha\beta}(-\omega)\, \ta_{\beta}\bigl(\omega-i\tfrac{\kap}{2}\bigr)^\dag
\ket{\vr^{1/2}}\bra{\vr^{1/2}}
\ta_\alpha\bigl(\omega-i\tfrac{\kap}{2}\bigr)\\
&= c \cdot 2\cos\bigl(\tfrac{\pi\kap}{2}\bigr)
\int\frac{d\omega}{2\pi} \sum_{\alpha,\beta}
\tXi^{\R}_{\alpha\beta}(-\omega)\,
\bket{\vr^{1/4}\tpsi_\beta\bigl(-\omega-i\tfrac{\kap}{2}\bigr)\vr^{1/4}}
\bbra{\vr^{1/4} \tpsi_\alpha\bigl(\omega-i\tfrac{\kap}{2}\bigr) \vr^{1/4}}.
\end{aligned}
\end{equation}
Note that the last equation implies that $\tXi^{\R}(-\omega)\geq 0$.

As a curiosity, let us also represent $\delta R$ as $c T \ket{\vr^{1/2}}\bra{\vr^{1/2}}$, where $T$ is a superoperator acting on the left (physical) subsystem:\footnote{In the discussion of $T$, the notation $X\cdot Y$ is understood as the left-right action on $\ket{\vr^{1/2}}\bra{\vr^{1/2}}$, namely, $X$ acts on the ket and $Y$ acts on the bra. This is in contrast with the previous usage of $X\bdot Y$, where $X$ acts on the left subsystem and $Y$ acts on the right subsystem. As we have clarified, $T$ acts on the left subsystem only.}
\begin{equation}
T= i \sin\bigl(\tfrac{\pi\kap}{2}\bigr) (P \cdot I-I \cdot P)
+ 2 \cos\bigl(\tfrac{\pi\kap}{2}\bigr) \calL,
\end{equation}
where $P$ is a Hermitian operator,
\begin{equation}
P= \int \frac{d\omega}{2\pi}\, \sinh(\pi\omega) \sum_{\alpha\beta}
\tXi^{\R}_{\alpha\beta}(\omega)\,
\tpsi_\alpha\bigl(-\omega-i\tfrac{\kap}{2}\bigr)
\tpsi_\beta\bigl(\omega-i\tfrac{\kap}{2}\bigr)
\end{equation}
and $\calL$ is a Lindbladian,
\begin{equation}
\calL = \int \frac{d\omega}{2\pi} e^{\pi \omega}
\sum_{\alpha,\beta} \tXi^{\R}_{\alpha\beta}
\begin{aligned}[t]
&(\omega)\Bigl(-i \tpsi_\beta\bigl(\omega-i\tfrac{\kap}{2}\bigr)
\cdot  \tpsi_\alpha\bigl(-\omega-i\tfrac{\kap}{2}\bigr) \\
& - \tfrac{1}{2} \tpsi_\alpha\bigl(-\omega-i\tfrac{\kap}{2}\bigr)
\tpsi_\beta\bigl(\omega-i\tfrac{\kap}{2}) \cdot  I 
- \tfrac{1}{2} I \cdot  \tpsi_\alpha\bigl(-\omega-i\tfrac{\kap}{2}\bigr)
\tpsi_\beta\bigl(\omega-i\tfrac{\kap}{2}\bigr) \Bigr).
\end{aligned}
\end{equation}
An interesting observation is that the Lindbladian part in $T$ comes with the prefactor of $\cos\frac{\pi\kap}{2}$, which is called the {\it decoherence factor} in \cite{KS17-soft}. In the limit of maximal chaos, $\kap\rightarrow 1$, it tends to zero.\footnote{This does not mean that the incoherent part in $\delta R$ vanishes, as the coefficient $c$ may diverge while the combination $c \cos\frac{\pi\kap}{2}$ approaching a finite limit.} 

\subsection{Size operator}\label{sec:size_operator}

A ``size operator'' is any positive-semidefinite operator that measures the magnitude of the TFD deformation. Since we consider a one-parameter family of quantum states, and only to the first order in the parameter $c$ (as defined by Eq.~\eqref{eq:pertcorr}), there are many ways to measure. This is based on an underlying assumption: the measurement takes place around a certain time $t_0$, say, $t_0=0$, whereas the deformation is produced by a relatively weak source at a distant time. Within the naive model, we may use the operator
\(
Q_{\R}^{(\text{naive})}
=\int \frac{d\omega}{2\pi} \frac{d\omega'}{2\pi}\sum_{\alpha,\beta}
v_{\alpha\beta}(\omega,\omega')\, \ta_{\alpha}\bigl(\omega+i\frac{\varkappa}{2}\bigr)^\dag \ta_{\beta}\bigl(\omega'+i\frac{\varkappa}{2}\bigr)
\)
with some positive-semidefinite $v$. To assure convergence (in view of the exponential factors in the definition~\eqref{naivea} of $\ta_\alpha(\omega)$), we may require that $v_{\alpha\beta}(\omega,\omega')$ decay sufficiently fast as $\omega,\omega'\to\infty$. In the time domain, the previous equation becomes
\begin{equation}\label{Qnaive}
Q_{\R}^{(\text{naive})}
=\int dt\,dt'\sum_{\alpha,\beta}
u_{\alpha\beta}(t,t')\,e^{-\kap(t+t')/2}
a(\chi_{\alpha}(it))^{\dag} a(\chi_{\beta}(it')),\qquad
u\geq 0,
\end{equation}
where, according to the definition~\eqref{Defa},
\begin{equation}
a(\chi_{\alpha}(it))
=\chi_\alpha\bigl(-\tfrac{\pi}{2}+it\bigr)\bdot I
+i I\bdot \chi_\alpha\bigl(\tfrac{\pi}{2}+it\bigr).
\end{equation}
The ``window function'' $u$ can be, for example, Gaussian to eliminate any potential divergence. (In practice, it is not an issue.) In the SYK case, the indices $\alpha,\beta$ are redundant because the naive model consists of a single field, $\chi(it)=\chi_j(it)$ for some fixed $j$.

In the actual SYK model, the TFD deformation affects all Majorana modes equally, and therefore, the size operator may be averaged over the modes. Changing the overall normalization, we arrive at the following definition:\footnote{For infinite temperature and $u_\R(t,t')=\delta(t)\delta(t')$, we get $Q_{\R}=\sum_{j}a(\chi_j)^{\dag} a(\chi_j)$, where $a(\chi_j)=\chi_j\bdot I+iI\bdot\chi_j$. In this case, $\bra{\calO}Q_{\R}\ket{\calO}$ is the ``size'' of the operator $\calO$ in the sense of Ref.~\cite{QiSt18}. See Ref.~\cite{Lucas:2018wsc,Mousatov:2019xmc,Lensky:2020ubw} for more proposals of operator size at finite temperature.} 
\begin{equation}\label{Qfull}
Q_{\R}=\sum_{j=1}^{N} \int dt\,dt'\,
u_{\R}(t,t')\,e^{-\kap(t+t')/2}
a(\chi_j(it))^{\dag} a(\chi_j(it')),\qquad
u_{\R}\geq 0.
\end{equation}
By construction, $Q_\R$ is positive-semidefinite, annihilates $|\rho^{\frac{1}{2}}\rangle$, and measures the magnitude of the forward-propagating scrambling mode. The definition of the operator $Q_\A$ measuring the backward-propagating scrambling mode is similar but involves the coefficient function $u_{\A}(t,t')e^{\kap(t+t')/2}$.

To elaborate a bit, the ``magnitude'' is understood as the coefficient $c$. However, it is convenient to normalize $Q_\R$ by the condition
\begin{equation}
\delta\bcorr{Q_{\R}}=Cc.
\end{equation}
Let us express it more explicitly. For each filed $\chi=\chi_j$, we have
\begin{equation}
\begin{aligned}
\delta\bcorr{a(\chi(it))^{\dag} a(\chi(it'))}
&=\delta\bcorr{i\chi\bigl(\tfrac{\pi}{2}+it\bigr)
\bdot \chi\bigl(\tfrac{\pi}{2}+it'\bigr)
+i\chi\bigl(-\tfrac{\pi}{2}+it'\bigr)
\bdot \chi\bigl(-\tfrac{\pi}{2}+it\bigr)}\\[2pt]
&=c\,2\cos\bigl(\tfrac{\pi\kap}{2}\bigr)e^{\kap(t+t')/2}
\VF^\R(\pi+i(t-t')).
\end{aligned}
\end{equation}
Thus, the function $u_\R$ in~\eqref{Qfull} is normalized as follows:
\begin{equation}
\int dt\,dt' u_\R(t,t')\,
\VF^\R(\pi+i(t-t')) = \frac{C}{2\cos(\frac{\pi\kap}{2})\,N}
=k_\R'(-\kap)\,(\VF^\A,\VF^\R).
\end{equation}
The last equality is a result from Ref.~\cite{GuKi18}. It involves a kinetic coefficient $k_\R'(-\kap)$ called ``branching time'' and a certain inner product between the advanced and retarded vertex functions.

Finally, we remark that the coefficient $c$ is, actually, the average value of the true TFD deformation magnitude $z_\R$ that is measured by $C^{-1}Q_\R$. Formally, $z_\R$ is an eigenvalue of $C^{-1}Q_\R$, which is positive because $Q_\R$ is positive. More intuitively, $z_\R$ is a random variable that is generated by a quantum process, amplified to become essentially classical, and affecting all Majorana modes. If $z_\R$ is fixed, all terms in \eqref{Qfull} have the same expectation value $\frac{C}{N}z_\R$, while their fluctuations are independent. Since $N$ is large, the fluctuations are not important.

\section{Nonlinear theory}\label{sec:nonlinear}

In this section, we interpret the operator $e^{-zQ_{\R}}$ (where $z$ is some number) as a source of a backward-propagating scrambling mode that can be treated using mean-field theory. Likewise, the operator $e^{-zQ_{\A}}$ generates a forward-propagating mode of the given magnitude $z$, whose exact form is obtained by solving a certain equation on the double Keldysh contour. The operators in question have norm less or equal to $1$ if $\Re z\geq 0$ and vanish as $\Re z\to +\infty$. Furthermore, they are analytic in $z$, which implies the desired analytic properties of their matrix elements mentioned in section~\ref{sec:derivation}.

\subsection{Matrix element interpretation of \texorpdfstring{$F^\R$, $F^\A$, $h^\R$, $h^\A$}{FR,FA,hR,hA}}\label{sec:melem}

We now consider the perturbation created by a pair of operators beyond the linear order. Let us insert the operators at complex times $\theta_3=\frac{\pi}{2}+it_3$ and $\theta_4=-\frac{\pi}{2}+it_4$ with $t_3\approx t_4<0$ (as we did before) and probe the resulting forward-propagating mode with $e^{-zQ_\R }$ at times around $0$. The matrix element of the operator $e^{-zQ_\R}$ has an expression similar to~\eqref{OTOC_melem} and may be interpreted in a dual way, as a measure of the backward-propagating mode generated by that operator:
\begin{equation}
F^\A_{X_3,X_4}\bigl(z;\,\tfrac{\pi}{2}+it_3,\,-\tfrac{\pi}{2}+it_4\bigr)
=\bbra{\rho^{1/4}X_4(it_4)^\dag\rho^{1/4}} e^{-zQ_\R }
\bket{\rho^{1/4}X_3(it_3)\rho^{1/4}}.
\end{equation}
This function can be analytically continued in $t_3$, $t_4$, and a similar function arises from probing the perturbation created by $X_1(\theta_1)$, $X_2(\theta_2)$ with $e^{-zQ_{\A}}$:
\begin{equation}
\begin{aligned}
F^\A_{X_3,X_4}(z;\theta_3,\theta_4)
&= \bbra{\rho^{1/2}X_4(\theta_4)^\dag} e^{-zQ_\R }
\bket{\rho^{1/2}X_3(\theta_3)} &&
\text{for }\, \pi\geq\Re\theta_3\geq 0\geq\Re\theta_4\geq -\pi,\\[3pt]
F^\R_{X_1,X_2}(z;\theta_1,\theta_2)
&= \bbra{X_1(\theta_1)^\dag\rho^{1/2}} e^{-zQ_\A}
\bket{X_2(\theta_2)\rho^{1/2}} &&
\text{for }\, \pi\geq\Re\theta_1\geq 0\geq\Re\theta_2\geq -\pi.
\end{aligned}
\end{equation}
(It is also true that $F^\R_{X_1,X_2}(z;\theta_1,\theta_2) =\bbra{\rho^{1/2}X_2(\theta_2)^\dag} e^{-zQ_\A} \bket{\rho^{1/2}X_1(\theta_1)}$.)

The Taylor series
\begin{equation}
e^{-zQ_\A} = \sum_{m=0}^\infty \frac{(-z)^m}{m!}  Q_{\A}^m,\qquad
e^{-zQ_\R} = \sum_{m=0}^\infty \frac{(-z)^m}{m!}  Q_{\R}^m
\end{equation}
imply similar expansions for the functions $F^\R$, $F^\A$. They are given by Eq.~\eqref{FRA} with
\begin{equation}\label{Upsilon_vs_Q}
\begin{aligned}
\bigl(e^{-i\kap(\theta_1+\theta_2)/2}\bigr)^m\,
\VF_{X_1,X_2}^{\R,m}(\theta_{1}-\theta_{2})
&= \bbra{X_1(\theta_1)^\dag\rho^{1/2}} Q_\A^m
\bket{X_2(\theta_2)\rho^{1/2}},\\[3pt]
\bigl(e^{i\kap(\theta_3+\theta_4)/2}\bigr)^m\,
\VF_{X_3,X_4}^{\A,m}(\theta_{3}-\theta_{4})
&=  \bbra{\rho^{1/2}X_4(\theta_4)^\dag} Q_\R^m
\bket{\rho^{1/2}X_3(\theta_3)}.
\end{aligned}
\end{equation}

Following the convention of section~\ref{sec:size_operator}, we denote the eigenvalues of $C^{-1}Q_\A$, $C^{-1}Q_\R$ by $z_\A,z_\R$. Let us consider the eigenvalue decompositions
\begin{equation}
C^{-1}Q_\A=\int_{0}^{\infty}z_\A\Pi_\A(z_\A)\,dz_\A,\qquad
C^{-1}Q_\R=\int_{0}^{\infty}z_\R\Pi_\R(z_\R)\,dz_\R.
\end{equation}
The first of them, together with the first equation in~\eqref{Upsilon_vs_Q}, implies that
\begin{equation}
\VF_{X_1,X_2}^{\R,m}(\theta_{1}-\theta_{2})
=\int_{0}^{\infty}\bigl(e^{i\kap(\theta_1+\theta_2)/2}Cz_\A\bigr)^m
\bbra{X_1(\theta_1)^\dag\rho^{1/2}} \Pi_\A(z_\A)
\bket{X_2(\theta_2)\rho^{1/2}}\,dz_\A
\end{equation}
On the other hand, $\VF^{\R,m}$ is related to the inverse Laplace transforms of $F^\R$, that is, to the function $h^\R$ in \eqref{Laplace}, \eqref{moments}. Thus, we obtain the first equation below (the second one is similar):
\begin{equation}
\begin{aligned}
\bbra{X_1(\theta_1)^\dag\rho^{1/2}} \Pi_\A(z_\A)
\bket{X_2(\theta_2)\rho^{1/2}}
&= e^{i\kap(\theta_1+\theta_2)/2}C\,
h_{X_1,X_2}^{\R}\bigl( \underbrace{e^{i\kap(\theta_1+\theta_2)/2}Cz_\A}_{y_\A};
\theta_{1}-\theta_{2}\bigr),\\[2pt]
\bbra{\rho^{1/2}X_4(\theta_4)^\dag} \Pi_\R(z_\R)
\bket{\rho^{1/2}X_3(\theta_3)}
&= e^{-i\kap(\theta_3+\theta_4)/2}C\,
h_{X_3,X_4}^{\A}\bigl( \underbrace{e^{-i\kap(\theta_3+\theta_4)/2}Cz_\R}_{y_\R};
\theta_{3}-\theta_{4}\bigr).
\end{aligned}
\end{equation}
Note that we have reproduced the relation~\eqref{z_vs_y} between the magnitudes $z_\A,z_\R$ of scrambling modes and the random variables $y_\A,y_\R$ pertaining to their sources. If $y_\A,y_\R>0$ and $0\leq\tau\leq 2\pi$, then $h_{X^\dag,X}^{\R}(y_\A;\tau)\geq 0$ and $h_{X^\dag,X}^{\A}(y_\R;\tau)\geq 0$. Upon suitable normalization (as in equation \eqref{normalized_h}), these functions may be interpreted as the probability distributions of $y_\A,y_\R$.

\subsection{Kinetic equation on the double Keldysh contour}
\label{sec: keldysh}

Let us now focus on the SYK model,
\begin{equation}
H= i^\frac{q}{2} \sum_{1<j_1\cdots j_q<N} J_{j_1,...,j_q} \chi_{j_1}...\chi_{j_q}, \qquad \{\chi_j,\chi_k\}=\delta_{jk}, \qquad \overline{J^2_{j_1,...,j_q}} = \frac{(q-1)!}{N^{q-1}} J^2.
\end{equation}
The ambiguities in the definition of $Q_\A$ (and similarly, $Q_\R$) may be resolved as follows:\footnote{We could further set $t_0=0$ and assume that $|z|\ll 1$. Fixing $z$ is inconsequential because $F^\R(z;\theta_1,\theta_2)$ depends on $e^{-i\kap(\theta_1+\theta_2)/2}z$ and $\theta_1-\theta_2$. However, it is more convenient not to constrain $z$, but rather, consider the limit $t_0\to-\infty$.}
\begin{equation}
Q_\A=u_\A e^{\kap t_0}\sum_{j=1}^{N}a(\chi_j(it_0))^\dag a(\chi_j(it_0)),\qquad
u_\A=\frac{C}{2\cos(\frac{\pi\kap}{2})\VF^\A(\pi)\,N}
=k_\R'(-\kap)\,\frac{(\VF^\A,\VF^\R)}{\VF^\A(\pi)}.
\end{equation}
We treat $Q_\A$ using mean-field theory, and in particular, assume that the individual terms commute. Furthermore, we may put all instances of $a^\dag$ in front of $a$, which results in this approximation:
\begin{equation}\label{normal_order}
\begin{aligned}
e^{-zQ_\A} \approx 1
&+\frac{-zu_{\A}e^{\kap t_0}}{1!}\sum_{j=1}^{N}
a(\chi_j(it_0))^\dag a(\chi_j(it_0))\\
&+\frac{(-zu_{\A}e^{\kap t_0})^2}{2!}\sum_{j_1=1}^{N}\sum_{j_2=1}^{N}
a(\chi_{j_1}(it_0))^\dag a(\chi_{j_2}(it_0))^\dag
a(\chi_{j_2}(it_0)) a(\chi_{j_1}(it_0)) +\cdots
\end{aligned}
\end{equation}
It is justified because the commutators between operators with $j_1\not=j_2$ are relatively small. To cover the $j_1=j_2$ case, we need to make sure that each individual term is small, that is, $zu_{\A}e^{\kap t_0}\ll 1$. This condition is satisfied by fixing $z$ and taking $t_0$ to $-\infty$.

Our goal is compute $F^\R\bigl(z;\,\tfrac{\pi}{2}+it_1,\,-\tfrac{\pi}{2}+it_2\bigr) =\lim_{t_0\to-\infty}W(t_1,t_2)$, where
\begin{equation}
W(t_1,t_2)
=\frac{1}{N}\sum_{k=1}^{N}
\bbra{\rho^{1/4}\chi_k(it_1)\rho^{1/4}}
e^{-zQ_\A}
\bket{\rho^{1/4}\chi_k(it_2)\rho^{1/4}}
\end{equation}
with the operator $e^{-zQ_\A}$ on the right-hand side approximated using Eq.~\eqref{normal_order}. We will derive an integral equation for the function $W$. From now on, the averaging over $k$ will be implicit. Let us consider the Taylor expansion in $z$. The zeroth-order term is the Wightman function,
\begin{equation}
W^{(0)}(t_1,t_2)=\Tr\bigl(\rho^{1/2}\chi(it_1)\rho^{1/2}\chi(it_2)\bigr)
=\corr{\chi(\pi+it_1)\chi(it_2)},
\end{equation}
so we interpret $W$ as the Wightman function on a perturbed background. The first-order term is expressed using the formula~\eqref{a_vs_grcomm} for the action of $a(X)$ on states:
\begin{equation}
W^{(1)}(t_1,t_2) = -zu_{\A}e^{\kap t_0}\sum_{j=1}^{N}
\bcorr{[\chi(\pi+it_1),\chi_j(\pi+it_0)]_\gr\, [\chi_j(it_0),\chi(it_2)]_\gr}.
\end{equation}
The second-order term involves double commutators such as $[\chi_{j_2}(it_0),[\chi_{j_1}(it_0),\chi(it_2)]_\gr]_\gr$. To organize the calculation, we represent commutators by operator placement on the double Keldysh contour as shown below. The operator $\chi_j(it_0)$ can be placed on the upper~($u$) or lower~($d$) side of fold $2$, and $\chi_j(\pi+it_0)$ is similarly placed on fold $1$. The operators $Y_2$, $Y_1$ are located on those folds to the right, and $\TT_c$ denotes the contour ordering:
\begin{equation}
\begin{aligned}
[\chi_j(it_0),Y_2]_\gr
&=\TT_{\rm c}\bigl(\chi_j^d(it_0)-\chi_j^u(it_0)\bigr)Y_2\\[8pt]
[Y_1,\chi_j(\pi+it_0)]_\gr
&=\TT_{\rm c}Y_1\bigl(\chi_j^u(\pi+it_0)-\chi_j^d(\pi+it_0)\bigr)
\end{aligned}\quad
\begin{tikzpicture}[scale=0.5,baseline={([yshift=0pt]current bounding box.center)}]
\draw [->,>=stealth] (-100pt,0pt) -- (150pt,0pt) node[right]{\scriptsize  $\Re t =\Im\theta$};
\draw [->,>=stealth] (0pt, -130pt) -- (0pt,40pt) node[right]{\scriptsize $\Im t=-\Re\theta$};
\draw[thick,blue,far arrow] (-100pt,0pt)--(90pt,0pt);
\draw[thick,blue,far arrow] (90pt,-6pt)--(-100pt,-6pt);
\draw[thick,blue] (-100pt,-6pt)--(-100pt,-60pt);
\filldraw (90pt,-3pt) circle (2pt) node[below right]{\scriptsize fold $2$};
\node at (60pt,8pt) {\scriptsize $u$};
\node at (60pt,-14pt) {\scriptsize $d$};
\node at (60pt,-52pt) {\scriptsize $u$};
\node at (60pt,-74pt) {\scriptsize $d$};
\draw[thick,blue,far arrow] (-100pt,-60pt)--(90pt,-60pt);
\draw[thick,blue,far arrow] (90pt,-66pt)--(-100pt,-66pt);
\draw[thick,blue] (-100pt,-66pt)--(-100pt,-120pt);
\filldraw (90pt,-63pt) circle (2pt) node[below right]{\scriptsize fold $1$};
\filldraw[red] (-40pt,0pt) circle (2pt);
\filldraw[red] (-40pt,-6pt) circle (2pt)
  node[black,below]{\scriptsize $\theta=it_0$};
\filldraw[red] (-40pt,-60pt) circle (2pt);
\filldraw[red] (-40pt,-66pt) circle (2pt)
  node[black,below]{\hspace{-5pt}\scriptsize $\theta=\pi+it_0$};
\end{tikzpicture}
\end{equation}
Thus, the operator $e^{-zQ_\A}$ corresponds to the insertion of $e^{-I_{\rm pert}}$ into a contour-ordered product, where
\begin{equation}\label{Delta_I}
I_{\rm pert}=zu_{\A}e^{\kap t_0}\sum_{j=1}^{N}
\bigl(\chi_j^u(\pi+it_0)-\chi_j^d(\pi+it_0)\bigr)
\bigl(\chi_j^d(it_0)-\chi_j^u(it_0)\bigr).
\end{equation}
This is an addition to the SYK action on the double Keldysh contour. It affects the fermionic Green function in the same way as self-energy does, so we will simply modify the latter.

Mean-field equations on the double Keldysh contour have been considered in~\cite{AFI16} and, specifically for the SYK model, in appendix~C of~\cite{ZGK20}. Let us briefly review them and extract the relevant parts. The Green function $G$ and the self-energy $\Sigma$ are matrices in the fold ($1,2$) and the flavor ($u,d$) indices. By definition,
\begin{equation}
G^{ab}_{\alpha\beta}(t,t')
=-i\bcorr{\TT_{\rm c} \chi^a(\tau_\alpha+it)\,\chi^b(\tau_\beta+it')},\qquad
a,b=u,d,\quad \alpha,\beta=1,2,
\end{equation}
where $\tau_1$ and $\tau_2$ are fixed; we assume that $\tau_1=\pi$ and $\tau_2=0$. The Schwinger-Dyson equations have the usual form,
\begin{equation}
(G_0^{-1}-\Sigma)G =1 =G(G_0^{-1}-\Sigma),
\end{equation}
where
\begin{equation}
(G_0^{-1})^{ab}_{\alpha\beta}
=i\delta_{\alpha\beta}\delta^{ab}\xi^{a}\partial_t,\qquad
\xi^a=\begin{cases}1&\text{if }a=u,\\-1&\text{if }a=d.\end{cases}
\end{equation}
The self-energy for the unperturbed SYK model is
\begin{equation}\label{Sigma_ud}
\Sigma^{ab}_{\alpha\beta}(t,t')
=-iJ^2\xi^a\xi^b\bigl(iG^{ab}_{\alpha\beta}(t,t')\bigr)^{q-1},
\end{equation}
and the addition of a term $\TT_{\rm c} \chi^a(\tau_\alpha+it_0)\chi^b(\tau_\beta+it_0')$ to the action changes $\Sigma^{ab}_{\alpha\beta}(t,t')=-\Sigma^{ba}_{\beta\alpha}(t',t)$ by $-i\delta(t-t_0)\delta(t'-t_0')$.

It is customary to represent the flavor structure in the Keldysh basis, $\chi^{\pm}=\frac{1}{\sqrt{2}}(\chi^u\pm\chi^d)$, so that
\begin{equation}
G = \begin{pmatrix}
G^\K & G^\R \\
G^\A & 0 
\end{pmatrix},\qquad
\Sigma = \begin{pmatrix}
0 & \Sigma^{\A} \\ 
\Sigma^{\R} & \Sigma^{\K}
\end{pmatrix}, \qquad 
G_0^{-1} = \begin{pmatrix}
0 & i\partial_t \\
i\partial_t  & 0
\end{pmatrix}.
\end{equation}
Here, each matrix element is itself a matrix in the fold index. The retarded and advanced Green functions, $G^\R$ and $G^\A$, are fold-diagonal, while the Keldysh function $G^\K$ has both diagonal and off-diagonal parts. In the Keldysh basis, equation~\eqref{Sigma_ud} becomes:
\begin{equation}\label{Sigma_Keldysh}
\begin{aligned}
i\Sigma^{\R}_{\alpha\beta}(t,t') &= J^2
\Bigl(\bigl(iG^{\K+\R}_{\alpha\beta}(t,t')/2\bigr)^{q-1}
-\bigl(iG^{\K-\R}_{\alpha\beta}(t,t')/2\bigr)^{q-1}\Bigr),\\[2pt]
i\Sigma^{\A}_{\alpha\beta}(t,t') &= J^2
\Bigl(\bigl(iG^{\K+\A}_{\alpha\beta}(t,t')/2\bigr)^{q-1}
-\bigl(iG^{\K-\A}_{\alpha\beta}(t,t')/2\bigr)^{q-1}\Bigr),\\[4pt]
i\Sigma^{\K}_{\alpha\beta}(t,t') &= J^2
\Bigl(\bigl(iG^{\K+\R-\A}_{\alpha\beta}(t,t')/2\bigr)^{q-1}
+\bigl(iG^{\K-\R+\A}_{\alpha\beta}(t,t')/2\bigr)^{q-1}\Bigr).
\end{aligned}
\end{equation}
(To derive the first two lines, the cases $t>t'$ and $t<t'$ have to be considered separately.)

We are interested in  $G^\K_{12}=-2iW$,\footnote{In Ref.~\cite{ZGK20}, the notation $G^\W=-W$ was used and the fold labels $1$ and $2$ were swapped, so the relation in question was $G^\K_{21}=2iG^\W$.} while the perturbation \eqref{Delta_I} changes $\Sigma^\K_{12}(t,t')=-\Sigma^\K_{21}(t',t)$ by $zu_{\A}e^{\kap t_0}\cdot 2i\delta(t-t_0)\delta(t'-t_0)$. Note that the fold-diagonal elements of $G$ and $\Sigma$ form a self-contained subsystem, and therefore, are as at thermal equilibrium. In particular, on both folds,
\begin{equation}
iG^\R(t,t')=-iG^\A(t',t)
=\theta(t-t')\,\corr{\{\chi(it),\chi(it')\}}.
\end{equation}
The relevant equations are $(i\partial_t-\Sigma^\R)G^\K -\Sigma^\K G^\A =0$ and $G^\R(i\partial_t-\Sigma^\R) =1$, which imply that $G^{\K}=G^{\R}\Sigma^{\K}G^{\A}$. To find $G^\K_{12}$, we take the unperturbed $\Sigma^\K_{12}$ from the last equation in~\eqref{Sigma_Keldysh} or obtain it directly from~\eqref{Sigma_ud}, and add the perturbation term to it. The result is as follows:
\begin{empheq}[box=\widebox]{align}
\label{SD1}
W(t_1,t_2) &=\int_{-\infty}^{+\infty} dt \int_{-\infty}^{+\infty}  dt'\,
G^\R(t_1,t) \Sigma^\W (t,t') G^\A (t',t_2),\\[3pt]
\label{SD2}
\Sigma^\W(t,t') &= J^2 W(t,t')^{q-1}
- zu_{\A}e^{\kap t_0}\delta(t-t_0)\delta(t'-t_0).
\end{empheq}
The initial conditions are defined by the thermal equilibrium; more exactly,
\begin{equation}
W(t_1,t_2)=\corr{\chi(\pi+it_1)\chi(it_2)}\quad\;
\text{if }t_1<t_0 \text{ or } t_2<t_0.
\end{equation}

The linearized version of equations \eqref{SD1}, \eqref{SD2} was considered in Refs.~\cite{MSW17,GuKi18}. Its solution, or the solution of the above equations to the first order in $z$, gives the early-time retarded OTOC:
\begin{equation}\label{OTOCR}
\begin{aligned}
W(t_1,t_2)&\approx \corr{\chi(\pi+it_1)\chi(it_2)}\\
&-zu_{\A}e^{\kap t_0}\frac{1}{N} \sum_{j,k} \theta(t_1-t_0) \theta(t_2-t_0)\,
\bcorr{\{\chi_j(\pi+it_1),\chi_k(\pi+it_0)\}
\{\chi_j(it_2),\chi_k(it_0)\}}.
\end{aligned}
\end{equation}
In general, equations \eqref{SD1}, \eqref{SD2} should be solved numerically. However, there are certain special cases where the equilibrium Green functions $G^\R$, $G^\A$ have a simple form such that the integral equation \eqref{SD1} can be transformed to a differential equation and analytically solved. In the following section and appendix~\ref{BrownianSYK}, we will show two such examples: the large-$q$ SYK and the Brownian SYK models.

\section{Example: the large-$q$ SYK model}\label{sec:large_q}

\subsection{Preliminaries}

The large-$q$ SYK model was introduced by Maldacena and Stanford, who computed, among other things, the two-point function and the Lyapunov exponent~\cite{MS16-remarks}. The four-point function was calculated in~\cite{Streicher19} using the mean-field approximation, which captures the early-time OTOC. We will compute the OTOC in the general case, $C^{-1}e^{\kap t}\sim 1$.

By definition, $N$ is taken to infinity before the $q\to\infty$ limit. To obtain sensible results, the coupling parameter $J$ should scale as follows:
\begin{equation}
J^2 = \frac{2^{q-1}}{q}\calJ^2,
\end{equation}
where $\calJ$ is fixed. The factor $2^{q-1}$ is to compensate the adopted normalization of Majorana operators, $\chi^2=1/2$. The $1/q$ factor in $J^2$ implies that the equilibrium self-energy is proportional to $1/q$. We will arrange that $u_\R=u_\A=2/q$ so that the perturbation scales in the same way. 

Instead of $q$ and $\calJ$, it is often convenient to use $\Delta$ and $v$ that are defined as follows:
\begin{equation}
\Delta=\frac{1}{q},\qquad\quad
\frac{v}{2\cos\frac{\pi v}{2}}=\calJ,\quad 0<v<1.
\end{equation}
The number $v$ determines the natural time scale in the system. For example, the Lyapunov exponent is $\kap=v$ and the two-point function is as follows:
\begin{equation}\label{G_large-q}
G(\theta_1,\theta_2)=\corr{\chi(\theta_1)\chi(\theta_2)}
=\frac{1}{2} 
\left( \frac{\cos\frac{v\pi}{2}}{\cos\frac{v(\pi-\theta_1+\theta_2)}{2}}
\right)^{2\Delta},\qquad
2\pi \geq \Re(\theta_1-\theta_2) \geq 0.
\end{equation}

Note one subtlety about the derivation of equation~\eqref{G_large-q}. The method of~\cite{MS16-remarks}, which we will also use, is based on the ansatz $G(\theta_1,\theta_2)=\frac{1}{2}e^{g(\theta_1,\theta_2)/q}$ and the assumption $|g(\theta_1,\theta_2)|\ll q$. This assumption is satisfied in the region where the self-energy is non-negligible. Indeed $\Sigma=J^{2}G^{q-1} =\frac{\calJ^2}{q}(e^{g/q})^{q-1} \approx\frac{\calJ^2}{q}e^{g}$. We have $|\Im g|\leq \frac{v\pi}{2}$; the real part of $g$ is always negative, and if it is large in magnitude, then $\Sigma$ is small. More specifically, the self-energy and all its variants are concentrated in the region $v|\Im(\theta_1-\theta_2)|\lesssim 1$. However $G$ and other correlation functions extend farther out. To verify the extended solution, one can first check that the retarded and advanced Green functions,
\begin{equation}\label{GRA_approx}
iG^\R(t_1,t_2)=-iG^\A(t_2,t_1)
\approx \theta(t_1-t_2)\,e^{-v(t_1-t_2)/q},
\end{equation}
satisfy the Schwinger-Dyson equation. Using the spectral function $A=i(G^\R-G^\A)$, one confirms that $G(\tau_1+it_1,\tau_2+it_2) \approx\frac{1}{2} e^{-v|t_1-t_2|/q}$.

These are some useful formulas pertaining to the early-time OTOC:
\begin{gather}
\kap=v,\qquad C = N\cdot 4\Delta^2\cos\frac{\pi v}{2},\qquad
\VF^\R(\theta) = \VF^\A(\theta) 
= \frac{\Delta}{\cos\frac{v(\pi-\theta)}{2}},\\[3pt]
k'(-\kap)=\frac{3}{2v},\qquad (\VF^\A, \VF^\R)=4\Delta^2\frac{v}{3}.
\end{gather}
We took them from~\cite{GuKi18} but changed the normalization of $C$, $\VF^\R$, $\VF^\A$ (while preserving the product $C^{-1}\VF^\R\VF^\A$) so that $u_\A=u_\R=2\Delta$.

\subsection{Differential form and solution of the kinetic equation}

Let us rewrite equations~\eqref{SD1}, \eqref{SD2} using the ansatz $W=\frac{1}{2}e^{g/q}$:
\begin{equation}
\label{eg_int_eq}
W(t_1,t_2)=\frac{e^{g(t_1,t_2)/q}}{2}
=\int_{-\infty}^{+\infty} dt \int_{-\infty}^{+\infty} dt'\,
G^\R(t_1,t)  \biggl(\frac{\calJ^2}{q}e^{g(t,t')}
-\frac{2z e^{\kap t_0}}{q} \delta(t-t_0)\delta(t'-t_0)\biggr) G^\A(t',t_2).
\end{equation}
Taking the derivatives with respect to $t_1$ and $t_2$, assuming that $|g(t_1,t_2)|\ll q$, and neglecting the $O(q^{-1})$ terms in $\partial_{1}G^\R(t_1,t)=-i\delta(t_1-t)+O(q^{-1})$ and $\partial_{2}G^\A(t',t_2)=i\delta(t_2-t')+O(q^{-1})$, we get the differential equation
\begin{equation}\label{Liouville}
\partial_1 \partial_2 \frac{g(t_1,t_2)}{2}
= \calJ^2 e^{g(t_1,t_2)} - 2z e^{\kap t_0} \delta(t_1-t_0) \delta (t_2-t_0).
\end{equation} 
It is a variant of the Liouville equation with an external source localized at $t_1=t_2=t_0$. A similar equation appears in the study of a quantum quench of the large-$q$ SYK model~\cite{EKSS17}.

\begin{figure}[t]
\center
\subfloat[Retarded solution]{
\begin{tikzpicture}[scale=0.6, baseline={(current bounding box.center)}]
\filldraw[white, fill=green!50] (0pt,0pt) rectangle (90pt,90pt);
\draw[->,>=stealth] (-100pt,0pt) -- (100pt,0pt) node[right]{$t_1-t_0$};
\draw[->,>=stealth] (0pt,-100pt) -- (00pt,100pt) node[above]{$t_2-t_0$};
\node at (20pt,20pt) {$A$};
\node at (-20pt,20pt) {$B$};
\node at (-20pt,-20pt) {$C$};
\node at (20pt,-20pt) {$D$};
\end{tikzpicture}}
\hspace{20pt}
\subfloat[Advanced solution]{
\begin{tikzpicture}[scale=0.6, baseline={(current bounding box.center)}]
\filldraw[white, fill=green!50] (0pt,0pt) rectangle (-90pt,-90pt);
\draw[->,>=stealth] (-100pt,0pt) -- (100pt,0pt) node[right]{$t_1-t_0$};
\draw[->,>=stealth] (0pt,-100pt) -- (00pt,100pt) node[above]{$t_2-t_0$};
\node at (20pt,20pt) {$A$};
\node at (-20pt,20pt) {$B$};
\node at (-20pt,-20pt) {$C$};
\node at (20pt,-20pt) {$D$};
\end{tikzpicture}}
\caption{Regions in the $(t_1,t_2)$ plane for equation~\eqref{Liouville}. The green color indicates the region that is affected by the source term. We focus on case (a), the retarded solution.}
\label{figRetAdv}
\end{figure}
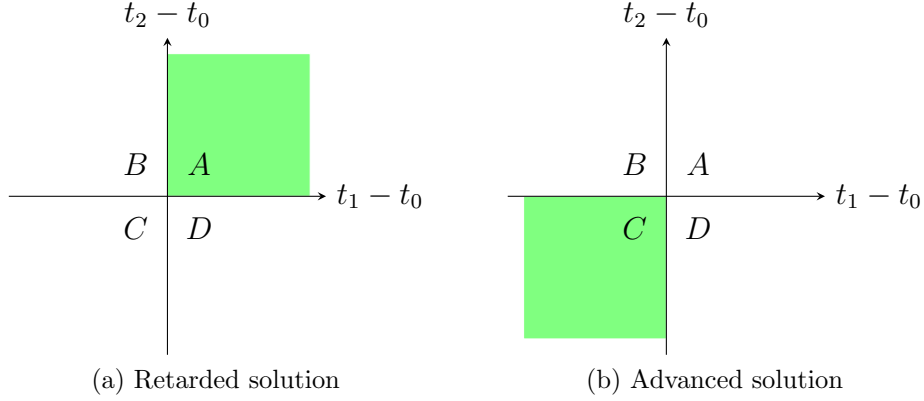

Equation~\eqref{Liouville} is a nonlinear wave equation with light cone coordinates $t_1$, $t_2$. As such, it has a retarded solution and an advanced solution as shown in Fig.~\ref{figRetAdv}; they correspond to $F^\R$ and $F^\A$, respectively. Here, we consider the retarded solution. It differs from the equilibrium solution $g^{(0)}$ (of the equation without the source) only in quadrant $A$, i.e.\ for $t_1\geq t_0$ and $t_2\geq t_0$. We expect $g$ to jump at the boundary of this region, but $\partial_1 g$ is continuous across the horizontal boundary part ($t_1>t_0$, $t_2=t_0$) and $\partial_2 g$ is continuous across the vertical boundary part ($t_1=t_0$, $t_2>t_0$). It follows that $g$ has a constant jump along the entire boundary:
\begin{equation}
g(t_1,t_2)=-4ze^{\kap t_0}\theta(t_1-t_0)\theta(t_2-t_0)
+\text{regular part}.
\end{equation}
Thus, we have determined the boundary conditions for $g$ in the upper right quadrant:
\begin{equation}\label{bc_g}
g(t_1,t_0)=g^{(0)}(t_1,t_0)-4ze^{\kap t_0},\qquad
g(t_0,t_2)=g^{(0)}(t_0,t_2)-4ze^{\kap t_0},
\end{equation}
where $g^{(0)}(t_1,t_2)$ corresponds to $W^{(0)}(t_1,t_2)=G(\pi+it_1,it_2)$, i.e.
\begin{equation}
e^{g^{(0)}(t_1,t_2)} = \bigl(2W^{(0)}(t_1,t_2)\bigr)^{q}
=\left( \frac{ \cos\frac{\pi v}{2} }{ \cosh\frac{v(t_1-t_2)}{2} } \right)^{2}.
\end{equation}

The general solution of the Liouville equation~\eqref{Liouville} without the source has the following form~\cite{Tsutsumi80}:
\begin{equation}
e^{g(t_1,t_2)} = -\calJ^{-2} \frac{f_1'(t_1)f'_2(t_2)}{(f_1(t_1)-f_2(t_2))^2}\,,
\end{equation}
where the function $f_1$ and $f_2$ are arbitrary. (The expression on the right-hand side has a remarkable $\operatorname{SL}(2,\RR)$ symmetry: it is invariant under the transformation $f_k(t)\mapsto\frac{af_k(t)+b}{cf_k(t)+d}$ for $k=1,2$.) In particular, the function $g^{(0)}$ (with $\frac{v}{2\cos\frac{\pi v}{2}}=\calJ$) is given by
\begin{equation}
f_1^{(0)}(t)=\coth\frac{v(t-t_0)}{2}\,,\qquad
f_2^{(0)}(t)=\tanh\frac{v(t-t_0)}{2}\,.
\end{equation}
The solution satisfying the boundary conditions~\eqref{bc_g} is closely related:
\begin{equation}
\label{eqn: new f}
f_1(t)=e^{2ze^{vt_0}}\coth\frac{v(t-t_0)}{2}\,,\qquad
f_2(t)=e^{-2ze^{vt_0}}\tanh\frac{v(t-t_0)}{2}\,.
\end{equation}
Thus, we have found the non-equilibrium Wightman function for $t_1>0$ and $t_2>0$:
\begin{equation}
\label{W_solution}
W(t_1,t_2)
= \frac{1}{2} \left(\frac{\cos \frac{\pi v}{2}}
{e^{2ze^{v t_0}} \cosh \frac{v(t_1-t_0)}{2} \cosh \frac{v(t_2-t_0)}{2}
-e^{-2ze^{vt_0}} \sinh \frac{v(t_1-t_0)}{2} \sinh \frac{v(t_2-t_0)}{2}}
\right)^{2\Delta}.
\end{equation}

Similarly to the discussion of the equilibrium two-point function, the Liouville equation is applicable in the neighborhood of the diagonal where $|g(t_1,t_2)|\ll q$. To verify the solution~\eqref{W_solution} away from the diagonal, we have to use the original integral equation. The self energy (i.e.\ the expression in big parentheses in~\eqref{eg_int_eq}) is concentrated in the region $v|t_1-t_2|\lesssim 1$. Due to the special form of the retarded and advanced Green functions~\eqref{GRA_approx}, the solution tails in the complementary region have the following form:
\begin{equation}\label{od_tails}
W(t_1,t_2)\approx\begin{cases}
\frac{1}{2}e^{\Delta(g_2(t_2)-vt_1)} &\text{if } v(t_1-t_2)\gg 1,\\
\frac{1}{2}e^{\Delta(g_1(t_1)-vt_2)} &\text{if } v(t_2-t_1)\gg 1.
\end{cases}
\end{equation}
The function $W$ we have found has this property. Since the Liouville region and the tail regions overlap, the exact tails (i.e.\ $g_1,g_2$) for our $W$ and for the actual solution coincide. Therefore, our results are valid in the entire plane.

\subsection{Vertex functions and late-time OTOC}\label{sec:large-q_OTOC}

We calculate the OTOC using the general formula~\eqref{OTOC1} and the relations involved in its derivation. In terms of the kinetic equation, $F^\R\bigl(z;\, \tfrac{\pi}{2}+it_1,\,-\tfrac{\pi}{2}+it_2\bigr) =W(t_1,t_2)$ in the $t_0\to-\infty$ limit (that is, in the case where the source is infinitely small but its effect at given $t_1,t_2$ is finite). Thus,
\begin{equation}
F^\R\bigl(z;\, \tfrac{\pi}{2}+it_1,\,-\tfrac{\pi}{2}+it_2\bigr)
=\frac{1}{2} \left( \frac{\cos\frac{\pi v}{2}}
{\cosh\frac{v(t_1-t_2)}{2}
+ze^{v\frac{t_1+t_2}{2}}} \right)^{2\Delta}.
\end{equation}
The analytic continuation to complex times is conveniently written as follows:
\begin{equation}\label{qFR}
F^\R(z;\theta_1,\theta_2)
=G(\theta_1,\theta_2)\,
\biggl(1+\frac{z_{12}}{\vth_{12}}\biggr)^{-2\Delta},
\end{equation}
where
\begin{equation}
z_{12}=e^{-iv\frac{\theta_1+\theta_2}{2}}z,\qquad
\vth_{12}=\cos\frac{v(\pi-\theta_1+\theta_2)}{2}.
\end{equation}

Let us comment on analytic properties of the function $F^\R$. First, $F^\R\bigl(z;\, \tfrac{\pi}{2}+it_1,\,-\tfrac{\pi}{2}+it_2\bigr)$ is analytic in $z$ in the entire plane, except for a branch cut from $-\infty$ to $-(e^{-vt_1}+e^{-vt_2})/2$. More generally, if $\pi\geq\Re\theta_1\geq 0\geq\Re\theta_2\geq -\pi$, then $F^\R\bigl(z;\theta_1,\theta_2)$ is analytic in the region $|\arg z|<\bigl(1-\frac{v}{2}\bigr)\pi$. This is consistent with the abstract result based on the consideration of matrix elements, which asserts the analyticity in the right half-plane under the same conditions. For the large-$q$ SYK model, the matrix element argument can be strengthened as follows. The vector $\ket{\chi(\theta_2)\rho^{1/2}}$ is well-defined for $\frac{1-v}{2v}\pi>\Re\theta_2>-\frac{1+v}{2v}\pi$ (vs.\ $0\geq\Re\theta_2\geq -\pi$ in the general case) because $\braket{\chi(\theta_2)\rho^{1/2}}{\chi(\theta_2)\rho^{1/2}} =G(-\theta_2^*,\theta_2)$ is bounded for all $\theta_2$ in the indicated range. If we rather keep $\theta_1$, $\theta_2$ the same, the range of possible $\arg z$ is extended by $\frac{1-v}{2}\pi$ in both directions, matching the actual analyticity domain.

The vertex functions $\VF^{R,m}$ entering equation~\eqref{FRA} can be found by Taylor expanding $F^\R(z;\theta_1,\theta_2)$ in $z_{12}$: 
\begin{equation}
F^\R(z;\theta_1,\theta_2)
=\sum_{m=0}^{\infty}
\frac{(-z_{12})^m}{m!}\cdot 
\underbrace{
G(\theta_1,\theta_2)
\frac{\Gamma(2\Delta+m)}{\Gamma(2\Delta)}\vth_{12}^{-m}
}_{\VF^{\R,m}(\theta_1-\theta_2)}.
\end{equation}
Due to the analyticity and decay at infinity in the right half-plane (and slightly beyond), $F^\R$ can be represented as a Laplace transform in $z_{12}$:
\begin{equation}\label{resGW}
F^\R(z;\theta_1,\theta_2)
=\int_0^\infty h^\R(y;\theta_1-\theta_2)\exp(-z_{12}y)\,dy.
\end{equation}
For the specific function~\eqref{qFR}, the inverse Laplace transform is:
\begin{equation}
h^\R(y;\theta_1-\theta_2)
=G(\theta_1,\theta_2)\,\frac{\vth_{12}^{2\Delta}}{\Gamma(2\Delta)}\,
y^{2\Delta-1}\exp(-\vth_{12}y).
\end{equation}
Note that $\VF^{\A,m}(\theta)=\VF^{\R,m}(\theta)$ and $h^\A(y,t)= h^\R(y,t)$ due to the time reversal symmetry of the SYK model.

Now, we use our main formula~\eqref{OTOC1} and find the OTOC:
\begin{equation}
\begin{aligned}
\OTOC(\theta_1,\theta_2;\theta_3,\theta_4)
&= \int_0^{+\infty}\! dy_\A  \int_0^{+\infty}\! dy_\R\, e^{-\lambda  y_\A y_\R}
h^\R(y_\A,\theta_1-\theta_2)\, h^\A(y_\R,\theta_3-\theta_4) \\
& = G(\theta_1,\theta_2)G(\theta_3,\theta_4)\,
\biggl(\frac{\vth_{12}\vth_{34}}{\lambda}\biggr)^{2\Delta}
U\biggl(2\Delta,1;\,\frac{\vth_{12}\vth_{34}}{\lambda}\biggr),
\end{aligned}
\end{equation}
where
\begin{equation}
\lambda =C^{-1}e^{iv(\pi-\theta_1-\theta_2+\theta_3+\theta_4)},\qquad
C=N\cdot 4\Delta^2\cos\frac{\pi v}{2},
\end{equation}
and $U(a,b,z)$ is the confluent hypergeometric function. For the symmetric configuration \eqref{eqn: equal time}, the factor $\lambda$ is real, $\lambda = C^{-1} e^{v\frac{t_1+t_2-t_3-t_4}{2}}$.

The OTOC can also be written in the form \eqref{eq:nlOTOC}. In the large-$q$ SYK case, it amounts to the asymptotic expansion
\begin{equation}
x^{-2\Delta}U(2\Delta,1;x^{-1})
=\sum_{n=0}^{\infty}\frac{\Gamma(2\Delta+n)^2}{\Gamma(2\Delta)^2}
\frac{(-x)^n}{n!}.
\end{equation}

\section{Discussion}\label{sec:discussion}

Our study adds to the well-established relation between quantum chaos, information scrambling, and the instability of the thermofield double. In particular, we have described the production of scrambling modes by operator pairs using the functions $h^\A$ and $h^\R$. This is a quantum process even in the large-$N$ limit, though the dual picture (involving $F^\R$ and $F^\A$, respectively) is mean-field. The large-$N$ approximation also leads to a simple interaction form between counter-propagating modes, $a(y_\A,y_\R)=\exp\bigl(-C^{-1}e^{i\kap(\pi-\theta_1-\theta_2 +\theta_3+\theta_4)/2}y_{\A}y_{\R}\bigl)$, but again, $a(y_\A,y_\R)$ may be interpreted as a quantum scattering amplitude. These results support the idea of ``scramblon'' as a quantum object (essentially, a Bose field) and suggest the possibility of a special form of quantum mechanics describing scrambling, and maybe even gravity. It must be non-unitary because it is meant to provide a coarse-grained description, but it should not introduce an ``arrow of time''.

As a rather simple idea along these lines, one may use a variant of 't~Hooft's action for gravitational shocks~\cite{tH90} and represent the scattering amplitude as follows:
\begin{equation}\label{tH_integral}
a(y_\A,y_\R)=\frac{e^{i\kap\pi/2}C}{2\pi i}\int dz_{\R}\,dz_{\A}\,
\exp\Bigl(e^{i\kap\pi/2} \bigl( Cz_{\R}z_{\A}
-e^{-i\kap(\theta_1+\theta_2)/2}y_{\A}z_{\R}
-e^{i\kap(\theta_3+\theta_4)/2}y_{\R}z_{\A}\bigr)\Bigr).
\end{equation}
The integral is taken over a suitable surface in $\CC^2$, and the saddle point of the action (i.e.\ the exponent in the above expression) is given by~\eqref{z_vs_y}. Combining equations~\eqref{OTOC1} and~\eqref{tH_integral}, we get
\begin{equation}\label{SYY_integral}
\begin{aligned}
\OTOC_{X_1,X_2,X_3,X_4}(\theta_1,\theta_2,\theta_3,\theta_4)
\hspace{-5cm}&\\[2pt]
&=\frac{e^{i\kap\pi/2}C}{2\pi i}\int dz_{\R}\,dz_{\A}\,
\exp\bigl(e^{i\kap\pi/2}Cz_{\R}z_{\A}\bigr)\,
F^\R_{X_1,X_2}\bigl(e^{i\kap\pi/2}z_\R;\theta_1,\theta_2\bigr)\,
F^\A_{X_3,X_4}\bigl(e^{i\kap\pi/2}z_\A;\theta_3,\theta_4\bigr).
\end{aligned}
\end{equation}
This is, essentially, the effective model proposed by Stanford, Yang, and Yao, see Eq.~(2.6) in~\cite{SYY21}. In comparison, our equation~\eqref{OTOC1} is more similar to Eq.~(2.4) (or~(8) in the arXiv version) of~\cite{ShSt14}  because in both cases, the integral is taken over null energies running from $0$ to $\infty$.\medskip

An interesting extension of our results, leading to further questions, has to do with analytic properties of the function $F^\R$ and the spectral function. In the most general case, $F^\R(z;\theta_1,\theta_2)$ is analytic in the half-plane $\Re z>0$, provided $\pi\geq\theta_1\geq 0\geq\theta_2\geq-\pi$. The last condition can be relaxed by applying the argument we have used for the large-$q$ SYK model. Suppose that the spectral function decays as
\begin{equation}
A(\omega)\sim e^{-\tau_*|\omega|}\quad \text{for }\, \omega\to\infty
\end{equation}
or even faster. Then the two-point function $G(\theta_1,\theta_2)$ is well-defined for $2\pi+\tau_*>\Re(\theta_1-\theta_2)>-\tau_*$, and therefore, the states $\ket{X_1(\theta_1)^\dag\rho^{1/2}}$ and $\ket{X_2(\theta_2)\rho^{1/2}}$ have bounded norm for
\begin{equation}
\pi+\frac{\tau_*}{2}>\theta_1>-\frac{\tau_*}{2},\qquad
\frac{\tau_*}{2}>\theta_2>-\pi-\frac{\tau_*}{2}.
\end{equation}
If $\theta_1$ and $\theta_2$ are fixed and satisfy these inequalities, then $F^\R(z;\theta_1,\theta_2)$ is analytic in the half plane $\Re z>0$. But $F^\R$ actually depends on certain combinations of its variables: $F^\R(z;\theta_1,\theta_2) =f^\R(e^{-i\kap(\theta_1+\theta_2)/2}z;\theta_1-\theta_2)$ (see~\eqref{F_vs_f}). Hence, $F^\R(z;\frac{\pi}{2}+it_1,-\frac{\pi}{2}+it_2)$ is analytic in the region
\begin{equation}\label{argz_bound}
|\arg z|<\biggl(1+\kap\biggl(1+\frac{\tau_*}{\pi}\biggr)\biggr)\frac{\pi}{2}.
\end{equation}

On the other hand, $F^\R(z;\frac{\pi}{2}+it,-\frac{\pi}{2}+it)=f^\R(e^{\kap t}z;\pi)$ must have a singularity at some negative value of $x=e^{\kap t}z$. Let us first prove a weaker statement: as $x$ goes to $-\infty$, the function in question either diverges at some finite value of the variable or grows faster than any exponential. Indeed, $f^\R(x;\pi)=\int_{0}^{\infty}e^{-yx}h^\R(y,\pi)\,dy$ (see \eqref{Laplace}), where $h^\R(y,\pi)\geq 0$ (see remark at the end of section~\ref{sec:melem}). It follows that for any positive $y_*$ such that $h^\R(y_*,\pi)\not=0$, we have $f^\R(x;\pi)\geq ce^{-y_*x}$, where $c=h^\R(y_*,\pi)>0$. Such a $y_*$ can be arbitrarily large because $h^\R(y,\ldots)$ is analytic in a neighborhood of the positive real axis, and therefore, can vanish only at a discrete set of points.

To proceed, we will use the the integral equations~\eqref{SD1}, \eqref{SD2} for the function $W(t_1,t_2) =F^\R(z;\frac{\pi}{2}+it_1,-\frac{\pi}{2}+it_2)$, where $z<0$ is fixed. So far we have proved that $W(t,t)$ grows super-exponentially (namely, faster than $\exp(ae^{\kap t})$ for any $a$), but the goal is to show that it diverges at some finite $t$. Initially (i.e.\ at times around $t_0$), $W(t_1,t_2)$ varies at the characteristic time scale of the order of $1$. As larger times, it grows increasingly faster as a function of $\frac{t_1+t_2}{2}$, while the dependence on $t_1-t_2$ remains less sharp. In the regime where the effective growth exponent tends to infinity, we may assume that $W(t_1,t_2)\approx W\bigl(\frac{t_1+t_2}{2})$ and that $G^\R(t_1,t)\approx -i\theta(t_1-t)$,\, $G^\A(t',t_2)\approx i\theta(t_2-t')$. Thus, the $W\to\infty$ asymptotic behavior is described by a simple equation:
\begin{equation}
W(t)= 4J^2\int_{-\infty}^{\infty}(t-t')\,\theta(t-t')\,W(t')^{q-1}\,dt'.
\end{equation}
It can be reduced to the second-order differential equation $\frac{d^2}{dt^2}W=4J^2 W^{q-1}$, whose solution diverges as $W(t)\sim (t_*-t)^{-2/(q-2)}$. Note that the temperature does not matter here because we discuss short-time dynamics. This argument was inspired by the calculation of high-frequency noise at infinite temperature in Ref.~\cite{FIK11}.

Given that $F^\R(z;\frac{\pi}{2}+it_1,-\frac{\pi}{2}+it_2)$ is analytic in the region~\eqref{argz_bound} but singular at some negative real $z$, we conclude that $\tau_*\leq\pi(\kap^{-1}-1)$, or in dimensional units,
\begin{equation}\label{spfun_bound}
\wideboxed{
\tau_*\leq \frac{\pi}{\kap}-\frac{1}{2T}.
}
\end{equation}
This is a nontrivial relation because it connects high-frequency properties to the Lyapunov exponent, which is bounded by $2\pi T$. It was derived by a completely different method (not specific to SYK-like models) in~\cite{PCASA18} for $T=\infty$ and in~\cite{AvDy19} in general. In the maximal chaos case, we get $\tau_*=0$, that is, the spectral function has subexponential (for example, polynomial) decay. In general, and in particular for the $q=4$ SYK model, the bound~\eqref{spfun_bound} is not tight, see appendix~\ref{sec:num_tau}.\medskip

These are some questions we have come across:
\begin{enumerate}
\item How to include dynamics in 't~Hooft's action? This seems possible but not straightforward because the forward-propagating mode is described by the kinetic equation on the double Keldysh contour, and the backward-propagating mode has a similar but separate description. A special case of the effective model~\eqref{SYY_integral} was derived in~\cite{SYY21} from the $G\Sigma$ action on a certain contour. It would be interesting to generalize that derivation and track the origin of the factor $e^{i\kap\pi/2}$.

\item What is the best theoretical bound on the analyticity domain of $F^\R$? We have explained the fact that for the large-$q$ SYK model,  $F^\R(z;\frac{\pi}{2}+it_1,-\frac{\pi}{2}+it_2)$ is analytic in $z$ in the entire plane with a branch cut from $-\infty$ to $0$. Furthermore, the numerical results in appendix~\ref{sec:numerics} indicate that the same is true for $q=4,6$. Is this a general property?

\item How to compute $1/N$ corrections to the OTOC? Such corrections are due to Feynman diagrams with ladders joining not only at the ends, but also in the middle.
\end{enumerate}

\section*{Acknowledgments}

We thank Douglas Stanford and Juan Maldacena for useful comments. Yingfei Gu and Pengfei Zhang also thank Shunyu Yao for explaining the paper~\cite{SYY21} to them. Yingfei Gu is supported  by the Simons Foundation through the ``It from Qubit'' program.  Alexei Kitaev is supported by the Simons Foundation under grant 376205 and through the ``It from Qubit'' program, as well as by the Institute of Quantum Information and Matter, a NSF Frontier center funded in part by the Gordon and Betty Moore Foundation. Pengfei Zhang acknowledges support from the Walter Burke Institute for Theoretical Physics at Caltech.

\appendix

\section{Past and future horizon representations}
\label{sec: horizon rep}

For concreteness, we consider the fermionic case, namely, a set of Majorana fields $\chi_\alpha(t)$ that satisfy Wick's theorem and have spectral function $A$. The idea is to represent them in terms of ($1+1$)-d chiral Majorana fields ${\psi}_\beta(x,t)={\psi}_\beta (x-t)$ whose spectral function is 
\begin{equation}
A^{(0)}_{\alpha\beta} (t,t') = \{ {\psi}_{\alpha}(t), {\psi}_\beta (t') \} = \delta_{\alpha\beta} \delta (t-t'), \qquad A^{(0)}_{\alpha\beta} (\omega) = \delta_{\alpha\beta}.
\end{equation}
A general representation of this kind has the form 
\begin{equation}
\chi_\alpha (t) = \int \sum_\beta L_{\alpha\beta} (t,t') {\psi}_\beta (t') dt' , \qquad L_{\alpha\beta} (t,t') \in \RR,
\end{equation}
or 
\begin{equation}
\tchi_\alpha(\omega)
= \sum_\beta L_{\alpha\beta} (\omega) \tpsi_\beta (\omega), \qquad
L_{\alpha\beta}(\omega)^* = L_{\alpha\beta} (-\omega), 
\end{equation}
where 
\begin{equation}
L(\omega) L(\omega)^\dagger = A(\omega) \quad\; \text{for }\, \omega \in \RR.
\end{equation}
Let us now impose a causality condition, $L_{\alpha\beta} (t,t')=0$ if $t-t' < \const$. (One interpretation is that $\psi_\beta(t)$ is a field on the past horizon of a black hole, whereas $\chi_\alpha(t)$ is the same type of field at a fixed spatial location outside the horizon.) We will denote this representation by $L^\R(\omega)$; it admits an analytic continuation to the upper half plane. The future horizon representation $L^\A(\omega)$ is defined similarly. Thus,
\begin{gather}
\label{birk}
A(\omega) = L^\R(\omega) L^\R(\omega)^\dagger
= L^\A(\omega) L^{\A}(\omega)^\dagger \quad\; \text{for }\, \omega \in \RR,
\\[3pt]
L^{\R}(-\omega^*)^* = L^{\R}(\omega)\;\, \text{for }\Im\omega\geq 0,\qquad
L^{\A}(-\omega^*)^* = L^{\A}(\omega)\;\, \text{for }\Im\omega\leq 0.
\end{gather}

Equation \eqref{birk} is related to the Birkhoff factorization problem: Given a continuous matrix-valued function $M$ on the unit circle with $\det M(z)\neq 0$, find a decomposition
\begin{equation}
M(z)= L(z) \cdot R(z) \quad\; \text{for }\, |z|=1,
\end{equation}
where $L$ is analytic for $|z|\leq 1$ and $R$ is analytic for $|z|\geq 1$, and both functions are non-degenerate in their definition domains. If $M(z)>0$ on the unit circle, then there exists a solution such that 
\begin{equation}
R(z) = L(1/z^*)^*,
\end{equation}
and it is unique up to a unitary transformation~\cite{Shmulian53}.

\section{Brownian OTOC}\label{BrownianSYK}

Another example that can be solved analytically is the Brownian SYK model~\cite{SSS18}. The OTOC has been computed numerically (using the operator growth picture) in~\cite{SPQSC19} and analytically in~\cite{SYY21}. Here, we give a shorter derivation using our general method.

The Hamiltonian is time-dependent: 
\begin{equation}
H= i^\frac{q}{2} \sum_{1<j_1\cdots j_q<N} J_{j_1,...,j_q}(t) \chi_{j_1}...\chi_{j_q},\qquad \overline{J_{j_1,...,j_q}(t)J_{j_1,...,j_q}(t')} = \frac{(q-1)!}{N^{q-1}} J\delta(t-t').
\end{equation}
Since it involves white noise, the equilibrium state corresponds to $\beta=0$, and so our previous conventions are not applicable. \emph{In this section, we use exclusively real time $t$ and measure it in physical units.} The natural time scale is set by $J$ or a related parameter called ``quasiparticle decay rate'',
\begin{equation}
\Gamma = 2^{2-q}J.
\end{equation}
A previous study~\cite{ZGK20} has found the retarded Green function, $G^\R(t,0) =-i\theta(t) e^{-\Gamma t/2}$. Consequently, the equilibrium Wightman function is $W^{(0)}(t,0):=\corr{\chi(t)\chi(0)}= \frac{1}{2} e^{-\Gamma |t|/2}$. We characterize the early-time OTOC using some calculations from~\cite{ZGK20} and normalizing $\VF^\R$, $\VF^\A$ in a convenient way:
\begin{gather}
\kap=(q-2)\Gamma,\qquad C = N\cdot \frac{1}{2(q-2)^2},\qquad
\VF^\R(t) = \VF^\A(t) 
= \frac{1}{2(q-2)}e^{-\frac{(q-1)\Gamma}{2}|t|},\\[3pt]
k'(-\kap)=\frac{1}{(q-1)\Gamma},\qquad (\VF^\A, \VF^\R)=\frac{(q-1)\Gamma}{4(q-2)^2}.
\end{gather}

The equations \eqref{SD1}, \eqref{SD2} for $W(t_1,t_2)$ out of equilibrium take the form
\begin{equation}
    \begin{aligned}
        W(t_1,t_2) &=\int_{-\infty}^{+\infty} dt  \int_{-\infty}^{+\infty} dt'  \ G^\R(t_1,t) \Sigma^\W(t,t') G^\A(t',t_2), \\
    \Sigma^\W (t,t') & = J \delta (t-t') W(t,t')^{q-1}-\frac{z}{2(q-2)} e^{\varkappa t_0} \delta(t_1-t_0) \delta   (t_2-t_0).
    \end{aligned}
\end{equation}
The above integral equation can be transformed into a differential equation:
\begin{equation}
\left(\partial_1 + \frac{\Gamma}{2}\right)\left(\partial_2+\frac{\Gamma}{2}\right) W(t_1,t_2) =J\delta(t_1-t_2) W^{q-1} -\frac{z}{2(q-2)} e^{\varkappa t_0} \delta(t_1-t_0) \delta	 (t_2-t_0).
\label{brownian eqn}
\end{equation}

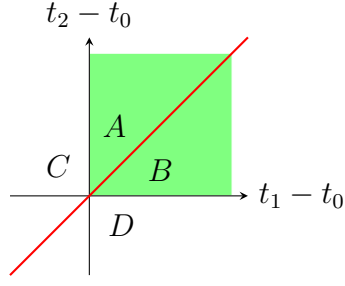
\begin{figure}[t]
\center
\begin{tikzpicture}[scale=0.6, baseline={(current bounding box.center)}]
\filldraw[white, fill=green!50] (0pt,0pt) rectangle (90pt,90pt);
\draw[->,>=stealth] (-50pt,0pt) -- (100pt,0pt) node[right]{$t_1-t_0$};
\draw[->,>=stealth] (0pt,-50pt) -- (00pt,100pt) node[above]{$t_2-t_0$};
\draw[red,thick] (-50pt,-50pt) -- (100pt,100pt);
\node at (30pt,30pt) [above left]{$A$};
\node at (30pt,30pt) [below right]{$B$};
\node at (-5pt,5pt) [above left]{$C$};
\node at (5pt,-5pt) [below right]{$D$};
\end{tikzpicture}
\caption{Regions for the retarded solution of equation~\eqref{brownian eqn}. The function $W(t_1,t_2)$ has a jump discontinuity at the boundary of the green-colored quadrant. In addition, the normal derivative is discontinuous at the red diagonal line.}
\label{fig:brownian}
\end{figure}

The retarded solution is divided into four regions as shown in  Fig.~\ref{fig:brownian}: 
\begin{itemize}
\item[A:] $t_2> t_1>t_0$. The general solution is given by 
\begin{equation}
W(t_1,t_2) = e^{- \frac{\Gamma}{2} t_1} f(t_2) + e^{- \frac{\Gamma}{2} t_2} g(t_1),
\end{equation}
where $f$ and $g$ are functions that will be fixed via boundary conditions. 
\item[B:] $t_1>t_2>t_0$. By reflection symmetry $t_1 \leftrightarrow t_2$, the general solution in this region can be related to region $A$:
\begin{equation}
W(t_1,t_2) = e^{- \frac{\Gamma}{2} t_1} g(t_2) + e^{- \frac{\Gamma}{2} t_2} f(t_1).
\end{equation}
\item[C:] $t_2>t_1$,\, $t_1<t_0$. This is a region with no influence of the perturbation source; therefore, the solution coincides with the equilibrium one,
$
W(t_1,t_2) = \frac{1}{2} e^{- \Gamma (t_2-t_1)/2} 
$.
\item[D:] $t_1>t_2$,\, $t_2<t_0$. Similarly to C, we have 
$
W(t_1,t_2) = \frac{1}{2} e^{- \Gamma (t_1-t_2)/2}
$. 
\end{itemize}
To fix the functions $f$ and $g$ in regions A and B, we match the jump of $W$ at the AC boundary and the jump of $\partial_{1}W$ at the AB boundary with the right-hand side of~\eqref{brownian eqn}. The final result is as follows (in regions A and B, i.e.\ for $t_1>t_0$ and $t_2>t_0$):
\begin{equation}
\quad W (t_1,t_2) = \frac{1}{2} e^{- \frac{\Gamma}{2} (t_1+t_2-2t_0)} \frac{1}{\left[e^{-(q-2) \Gamma (t_1+t_2-|t_1-t_2|-2t_0)/2} + (1-z e^{\varkappa t_0}/(q-2))^{2-q}-1 \right]^{\frac{1}{q-2}}}.
\end{equation}

Now, we obtain $F^\R$ as the $t_0\rightarrow -\infty$ limit of $W$ and then extract the vertex functions:
\begin{equation}
\begin{aligned}
F^\R(z  ;t_1,t_2) &=  \frac{1}{2} e^{- \frac{\Gamma}{2} (t_1+t_2)} \frac{1}{\left[e^{-\frac{\varkappa}{2}  (t_1+t_2-|t_1-t_2|)}  +z  \right]^{\frac{1}{q-2}}}   \\
&= \sum_{m=0}^{\infty} \frac{(-z_{12})^m}{m!} \cdot \underbrace{  \frac{1}{2} e^{- \frac{\Gamma}{2}|t_1-t_2|}  \frac{\Gamma(\frac{1}{q-2} +m)}{\Gamma(\frac{1}{q-2}) } e^{- \frac{m\varkappa}{2} |t_1-t_2|} }_{\VF^{\R,m}(t_{12})}.
\end{aligned}
\end{equation}
Here $z_{12}=z e^{\varkappa \frac{t_1+t_2}{2}}$. To obtain the late time OTOC, we write 
\begin{equation}
F^\R(z ;t_1,t_2) = \int_0^{+\infty} h^\R(y;t_1-t_2)\, e^{- y z_{12}}\, dy
\end{equation}
and find $h^\R$ via inverse Laplace transform:
\begin{equation}
h^\R(y;t) = \frac{y^{\frac{1}{q-2}-1}}{2 \Gamma(\frac{1}{q-2})}\,
e^{-y/a}  ,\qquad a= e^{- \frac{\varkappa}{2} |t|}.
\end{equation}
Due to the time reversal symmetry, $h^\A=h^\R$. Adapting equation~\eqref{OTOC1} to infinite temperature and real times, we get
\begin{equation}
\begin{aligned}
\OTOC(t_1,t_2;t_3,t_4)= \frac{1}{4  \lambda ^{\frac{1}{q-2}} }  U\left(\frac{1}{q-2},1, \frac{e^{\frac{\varkappa}{2} (|t_{12}|+|t_{34}|)}}{ \lambda }\right),\qquad
\lambda = C^{-1} e^{ \varkappa (t_1+t_2-t_3-t_4)/2 }.
\end{aligned}
\end{equation}

\section{Numerical study of the SYK model at finite $q$}\label{sec:numerics}

\subsection{Solution of the kinetic equation}\label{sec:num_kinetic}

In the main text and appendix \ref{BrownianSYK}, we have analytically derived $W(t_1,t_2)$ and the late-time OTOC for the large-$q$ static SYK model and the Brownian SYK model. In both cases, the integral equation for $W$ can be reduced to a differential equation, leading to a great simplification. In this appendix, we consider the static SYK model at finite $q$ by directly solving the integral equation.

\begin{figure}[t]
\center
\subfloat[Double Keldysh contour]{
\begin{tikzpicture}[scale=0.65,baseline={([yshift=0pt]current bounding box.center)}]
\draw [->,>=stealth] (-50pt,0pt) -- (200pt,0pt) node[right]{\scriptsize  $\Re(t)$};
\draw [->,>=stealth] (0pt, 20pt) -- (0pt,-150pt) node[right]{\scriptsize $\Im(t)$};
\draw[thick,blue,far arrow] (0pt,0pt)--(140pt,0pt);
\draw[thick,blue,far arrow] (140pt,-4pt)--(0pt,-4pt);
\draw[thick,blue] (0pt,-4pt)--(0pt,-60pt);
\filldraw (140pt,-2pt) circle (2pt) node[below right]{\scriptsize $t$};
\filldraw (140pt,-2pt) circle (2pt) node[above right]{\scriptsize fold 2};
\node at (100pt,5pt) {\scriptsize $u$};
\node at (100pt,-9pt) {\scriptsize $d$};
\node at (100pt,-55pt) {\scriptsize $u$};
\node at (100pt,-69pt) {\scriptsize $d$};
\draw[thick,blue,far arrow] (0pt,-60pt)--(140pt,-60pt);
\draw[thick,blue,far arrow] (140pt,-64pt)--(0pt,-64pt);
\draw[thick,blue] (0pt,-64pt)--(0pt,-120pt);
\filldraw (140pt,-62pt) circle (2pt) node[below right]{\scriptsize $\beta/2+3t$};
\filldraw (140pt,-62pt) circle (2pt) node[above right]{\scriptsize fold 1};
\filldraw (0pt,-120pt) circle (1pt) node[left]{\scriptsize $\beta+4t$};

\filldraw (0pt,-60pt) circle (1pt) node[above left]{\scriptsize $\beta/2+2t$};
\filldraw (0pt,-60pt) circle (1pt) node[below left]{\scriptsize $\beta/2+4t$};

\filldraw (0pt,0pt) circle (1pt) node[above left]{\scriptsize $0$};
\filldraw (0pt,-4pt) circle (1pt) node[below left]{\scriptsize $2t$};

\filldraw[red] (0pt,0pt) circle (1.5pt);
\filldraw[red] (0pt,-4pt) circle (1.5pt);

\filldraw[red] (0pt,-60pt) circle (1.5pt);
\filldraw[red] (0pt,-64pt) circle (1.5pt);
\end{tikzpicture}
}
\hspace{5pt}
\subfloat[$G(l,l')$]{
\begin{tikzpicture}[scale=0.3, baseline={(current bounding box.center)}]
        \draw (0,0) -- (10,0) -- (10,10) -- (0,10) -- (0,0);
        \draw[dashed] (2,0) -- (2,10);
        \draw[dashed] (4,0) -- (4,10);
        \draw[dashed] (5,0) -- (5,10);
        \draw[dashed] (7,0) -- (7,10);
        \draw[dashed] (9,0) -- (9,10);
        \draw[dashed] (0,1) -- (10,1);
        \draw[dashed] (0,3) -- (10,3);
        \draw[dashed] (0,5) -- (10,5);
        \draw[dashed] (0,6) -- (10,6);
        \draw[dashed] (0,8) -- (10,8);

        \node at (-2.5,9.5) {\scriptsize $l$};
        \node at (-2,9) {$\downarrow$};

        \node at (0.5,11.7) {\scriptsize $l'$};
        \node at (1,11) {$\rightarrow$};

        \filldraw (0,10) circle (0) node[left]{\scriptsize $0$};
        \filldraw (0,8) circle (0) node[left]{\scriptsize $t$};
        \filldraw (0,6) circle (0) node[left]{\scriptsize $2t$};
        \filldraw (0,5) circle (0) node[left]{\scriptsize $\frac{\beta}{2}+2t$};
        \filldraw (0,3) circle (0) node[left]{\scriptsize $\frac{\beta}{2}+3t$};
        \filldraw (0,1) circle (0) node[left]{\scriptsize $\frac{\beta}{2}+4t$};
        \filldraw (0,0) circle (0) node[left]{\scriptsize $\beta+4t$};
        \fill[pattern=north west lines, pattern color=blue] (0,5) rectangle (2,3);
\end{tikzpicture}
}
\hspace{20pt}
\subfloat[$f(l)f(l')$]{
\begin{tikzpicture}[scale=0.4, baseline={(current bounding box.center)}]
        \draw (0,0) -- (10,0) -- (10,10) -- (0,10) -- (0,0);

        \fill[gray!50] (0,10) rectangle (2,8);
        \fill[gray!50] (2,8) rectangle (4,6);
        \fill[gray!50] (5,10) rectangle (7,8);
        \fill[gray!50] (7,8) rectangle (9,6);
        \fill[gray!50] (5,5) rectangle (7,3);
        \fill[gray!50] (7,3) rectangle (9,1);
        \fill[gray!50] (0,5) rectangle (2,3);
        \fill[gray!50] (2,3) rectangle (4,1);

        \fill[red!50] (0,5) rectangle (2,6);
        \fill[red!50] (0,0) rectangle (2,1);
        \fill[red!50] (5,5) rectangle (7,6);
        \fill[red!50] (5,0) rectangle (7,1);

        \fill[red!50] (4,10) rectangle (5,8);
        \fill[red!50] (9,10) rectangle (10,8);
        \fill[red!50] (4,5) rectangle (5,3);
        \fill[red!50] (9,5) rectangle (10,3);

        \fill[blue!50] (2,5) rectangle (4,6);
        \fill[blue!50] (2,0) rectangle (4,1);
        \fill[blue!50] (7,5) rectangle (9,6);
        \fill[blue!50] (7,0) rectangle (9,1);

        \fill[blue!50] (4,8) rectangle (5,6);
        \fill[blue!50] (9,8) rectangle (10,6);
        \fill[blue!50] (4,3) rectangle (5,1);
        \fill[blue!50] (9,3) rectangle (10,1);
\end{tikzpicture}
}
\caption{(a) The double Keldysh contour used in the numerics is shown in blue; the red dots represent the insertion of sources. (b) The coordinate system for the Green function $G(l,l')$. The part that we use, $W(t_1,t_2)$, corresponds to the hatched area. (c) The function $f(l)f(l')$, where the white, gray, red, and blue colors represent $1$, $-1$, $i$, and $-i$, respectively.}
\label{fig:numerics1}
\end{figure}
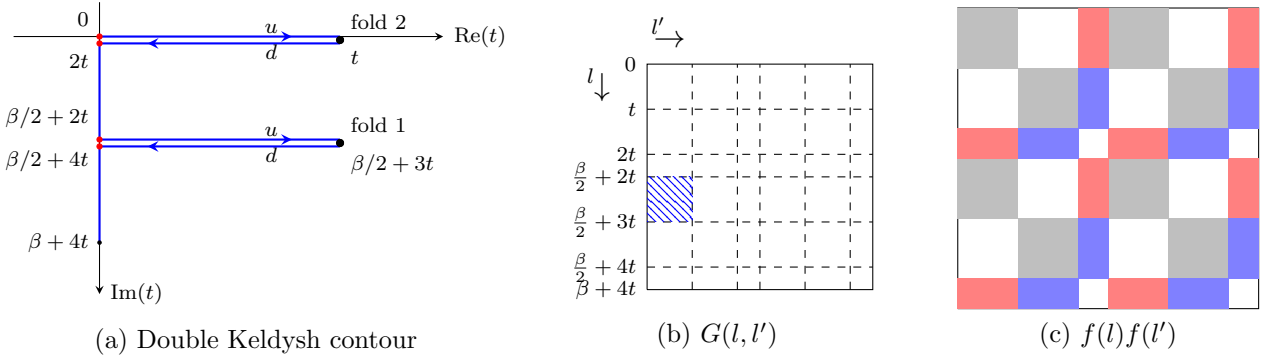

Instead of obtaining the retarded and advanced Green functions and using them in~\eqref{SD1} and~\eqref{SD2}, we directly solve the Schwinger-Dyson equation for the contour-ordered Green function on the full contour shown in figure \ref{fig:numerics1}\,(a). To simplify the notation, we consider time relative to $t_0$; this will be taken into account when interpreting the results. The contour is parametrized by a real parameter $l\in (0,4t+\beta)$. Here $l\in (0,2t)$ corresponds to the evolution in fold~2, where for $l\in (0,t)$ the system evolves forward in real time and for $l\in (t,2t)$ the system evolves backward in real time. Similarly, $l\in (2t+\beta/2,4t+\beta/2)$ corresponds to fold~1, which can be divided into the forward evolution part $l\in (2t+\beta/2,3t+\beta/2)$ and the backward evolution part $l\in (3t+\beta/2,4t+\beta/2)$. For $l\in (2t,2t+\beta/2)$ and $l \in (4t+\beta/2,4t+\beta)$, the system evolves in imaginary time, which connects folds 1 and~2. 

We first consider the problem without perturbation. The definition of the partition function reads
\begin{equation}
Z=\int D\chi_j~\exp\left(-\int_0^{\beta+4t}dl~\Big[\sum_j\frac{1}{2}\chi_j(l)\partial_l\chi_j(l)+f(l)H[\chi_j(l)]\Big]\right).
\end{equation}
Here, we have encoded the evolution direction into the function $f(l)$ defined as follows:
\begin{equation}
f(l) = 
\left\{\begin{array}{ll} 1,& l \in (2t,\frac{\beta}{2}+2t)\cup(\frac{\beta}{2}+4t,\beta+4t),\\
 i, & l \in (0,t)\cup(\frac{\beta}{2}+2t,\frac{\beta}{2}+3t),\\
 -i, & l \in (t,2t)\cup(\frac{\beta}{2}+3t,\frac{\beta}{2}+4t).
 \end{array}\right.
\end{equation}
The Schwinger-Dyson equation for $G(l,l')\equiv \frac{1}{N}\sum_j\left<\chi_j(l)\chi_j(l')\right>$ reads
\begin{equation}\label{eq:num1}
    \partial_lG(l,l') -\int_0^{\beta+4t} dl''~\Sigma(z=0,l,l'')G(l'',l')=\delta(l-l')
\end{equation}
with the self-energy 
\begin{equation}\label{eq:num2}
    \Sigma(z=0,l,l')=J^2f(l)f(l')G(l,l')^{q-1}.
\end{equation}

In our numerics, we discretize the continuous time $l$ into a lattice with spacing $\Delta l$. In practice, setting $\Delta l\sim 0.1J^{-1}$ already gives a good approximation for the Green function~\cite{Zhang:2020kia,Chen:2020wiq}. The contour contains $L=(\beta+4t)/\Delta l$ points. Thus, $G$ is represented by the $L\times L$ matrix with the elements $G_{mn}= G((m-1/2)\Delta l,(n-1/2)\Delta l)$, and $f(l)$ becomes $f_n= f((n-1/2)\Delta l)$. We now translate~\eqref{eq:num1} and~\eqref{eq:num2} into matrix equations:
\begin{equation}
G=(G_0^{-1}-\Sigma)^{-1}, \ \ \ \ \ \ \ (\Sigma(z=0))_{mn}=J^2\Delta l^2 f_m f_n G_{mn}^{q-1}.
\end{equation}
Note that instead of discretizing $\partial_l$, we directly use the inverse of the non-interacting Green function $(G_0)_{mn}=\frac{1}{2}\text{sgn}(m-n)$ to improve convergence. For nonzero $z$, there is an additional source added to the action (inserted at the red dots in figure \ref{fig:numerics1}\,(a)):
\begin{equation}
\begin{aligned}
\delta I=-s \sum_j
\Bigl(&\chi_j(0)\chi_j(2t+\beta/2)+\chi_j(2t)\chi_j(4t+\beta/2)\\[-8pt]
&-\chi_j(2t)\chi_j(2t+\beta/2)-\chi_j(0)\chi_j(4t+\beta/2)\Bigr).
\end{aligned}
\end{equation}
Here, we have defined $s=zu_Ae^{\kap t_0}$ for conciseness. After discretization, this leads to the modified Schwinger-Dyson equation
\begin{equation}\label{eq:num_matrix}
G=(G_0^{-1}-\Sigma)^{-1}, \ \ \ \ \ \ \ (\Sigma(z))_{mn}=J^2\Delta l^2 f_m f_n G_{mn}^{q-1}+s\sigma_{mn}
\end{equation}
with 
\begin{equation}
\sigma_{mn}=\left(\delta_{m,1}\delta_{n,\frac{2t+\beta/2}{\Delta l}}+\delta_{m,1+\frac{2t}{\Delta l}}\delta_{n,\frac{4t+\beta/2}{\Delta l}}-\delta_{m,1+\frac{2t}{\Delta l}}\delta_{n,\frac{2t+\beta/2}{\Delta l}}-\delta_{m,1}\delta_{n,\frac{4t+\beta/2}{\Delta l}}\right)-(m\leftrightarrow n).
\end{equation}

The numerical solution is obtained as follows. Using some initial guess for $G_{mn}$, we solve equation~\eqref{eq:num_matrix} iteratively until $G$ converges. The function $W$ representing the correlations between two folds can be extracted from the data. For this purpose, we can use either side ($u$ or $d$) of each fold; the $uu$ choice leads to the following equation, which is illustrated by figure~\ref{fig:numerics1}\,(b):
\begin{equation}
    W(t_1,t_2)=G(t_2+2t+\beta/2,t_1)\qquad \text{for}\ \ t_1,t_2\in (0,t).
\end{equation}

\begin{figure}[t]
    \centering
    \includegraphics[width=0.33\linewidth]{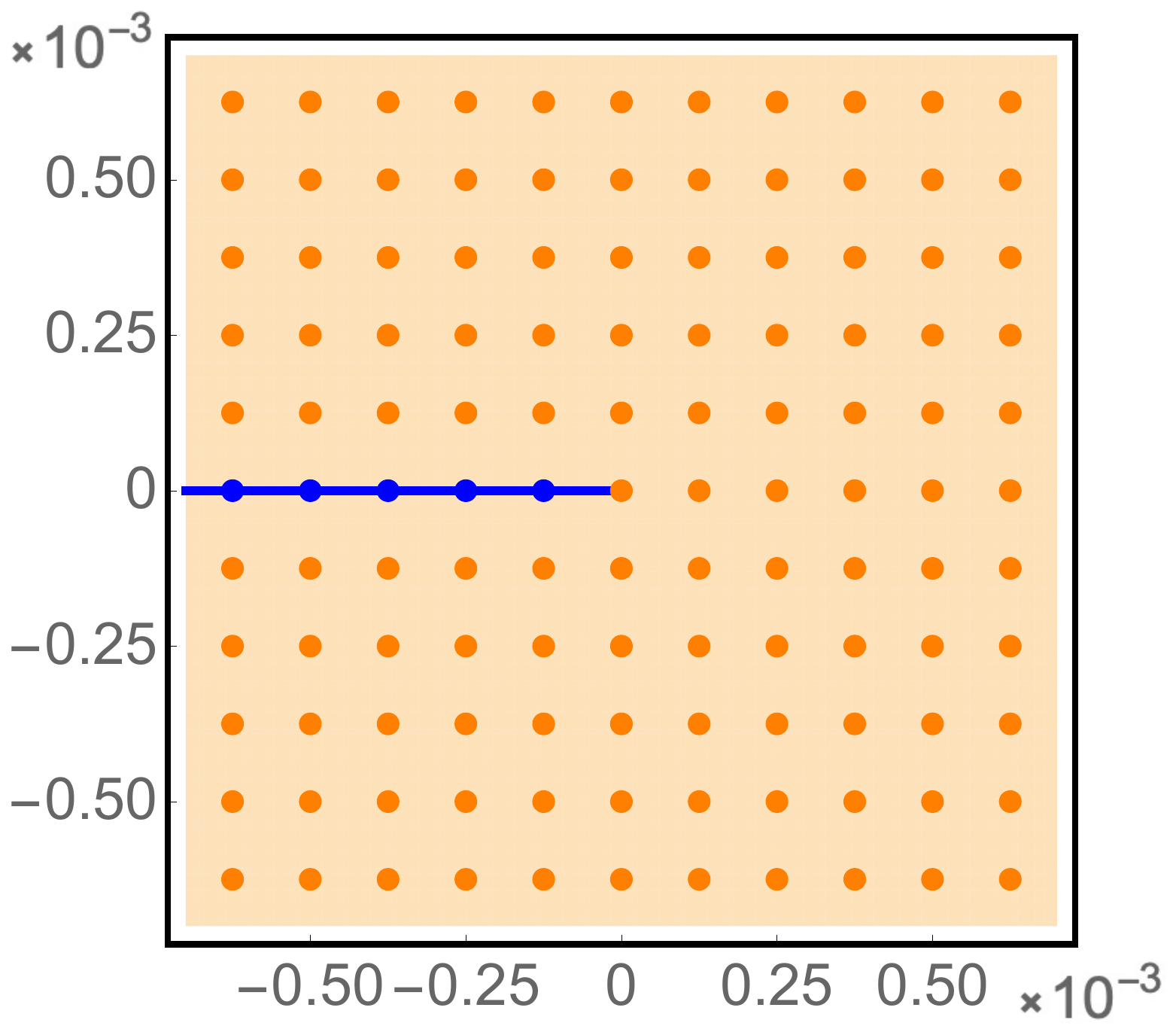}
    \caption{The convergence of $G$ on the complex plane of $s=zu_{\A}e^{\kap t_0}$. The orange/blue dots correspond to numerical tests where $G$ can/cannot converge, indicating whether the corresponding $W(t_1,t_2)$ is well-defined for all $t_1$, $t_2$. The result shows that $W$ is well-defined for $|\arg s|< \pi$, which serves as evidence that $F^\R(z;\frac{\pi}{2}+it_1,-\frac{\pi}{2}+it_2)$ is analytic for $|\arg z|< \pi$. This has been tested for $q=4,6$ with $\beta J=1,4,8,15$. }
    \label{fignum2}
\end{figure}

Ultimately, we are interested in the function $F^\R(z;\frac{\pi}{2}+it_1,-\frac{\pi}{2}+it_2)$, which is obtained as the $t_0\to-\infty$ limit of $W(t_1-t_0,t_2-t_0)$ with $s=zu_{\A}e^{\kap t_0}$. We expect that the numerical convergence of $W$ for $\arg s$ in a given interval and $|s|$ sufficiently small is equivalent to $F^\R(z;\frac{\pi}{2}+it_1,-\frac{\pi}{2}+it_2)$ being well-defined and analytic in $z$ in the same interval of $\arg z$ for all $t_1,t_2$. According to this criterion, our numerical results imply the analyticity of $F^\R$ in the interval $|\arg z|< \pi$, see figure~\ref{fignum2}. We have tested this for $q=4,6$ and $\beta J=1,4,8,15$.

\begin{figure}[p]
  \center
  \subfloat[The numerical result for $s=0.001$.]{\includegraphics[scale=0.4]{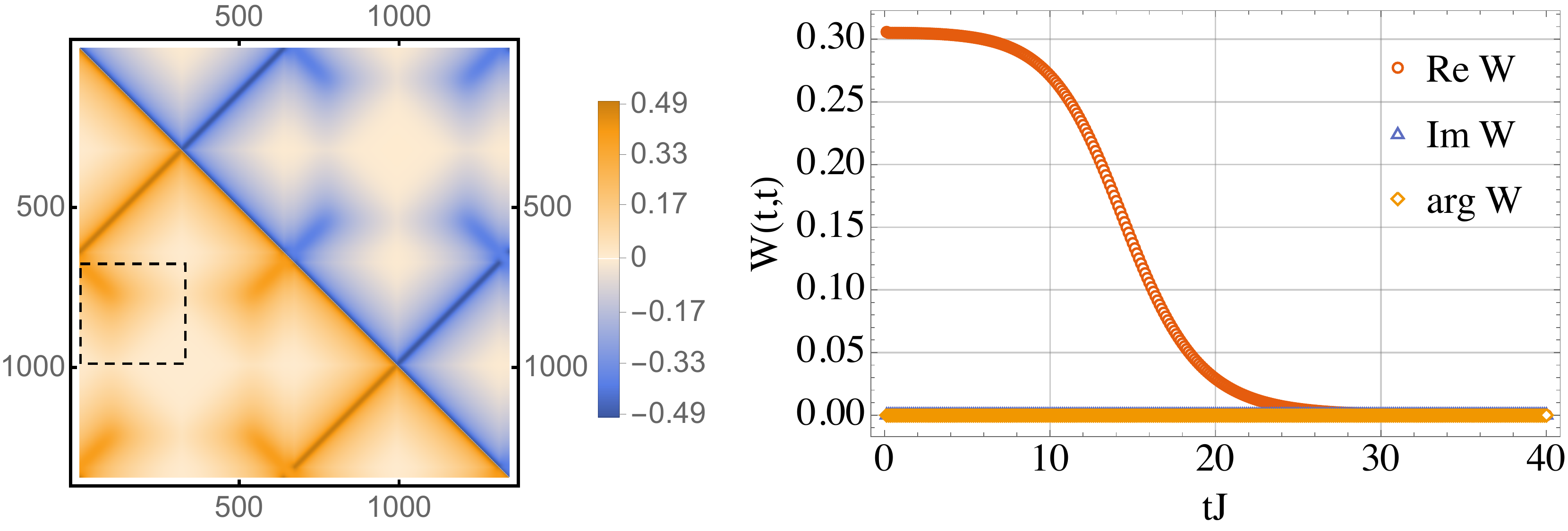}}\\
    \subfloat[The numerical result for $s=0.001e^{-\frac{\pi}{2} i}$.]{\includegraphics[scale=0.4]{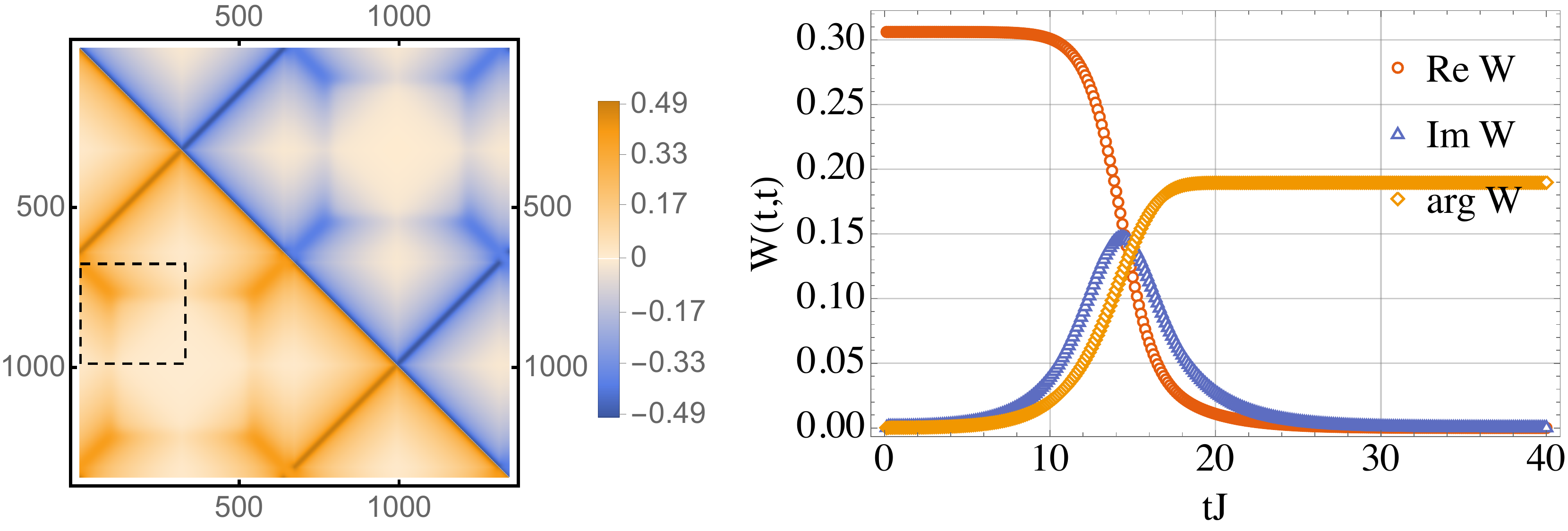}} \\
    \subfloat[The numerical result for $s=0.001e^{-\frac{3\pi}{4} i}$.]{\includegraphics[scale=0.4]{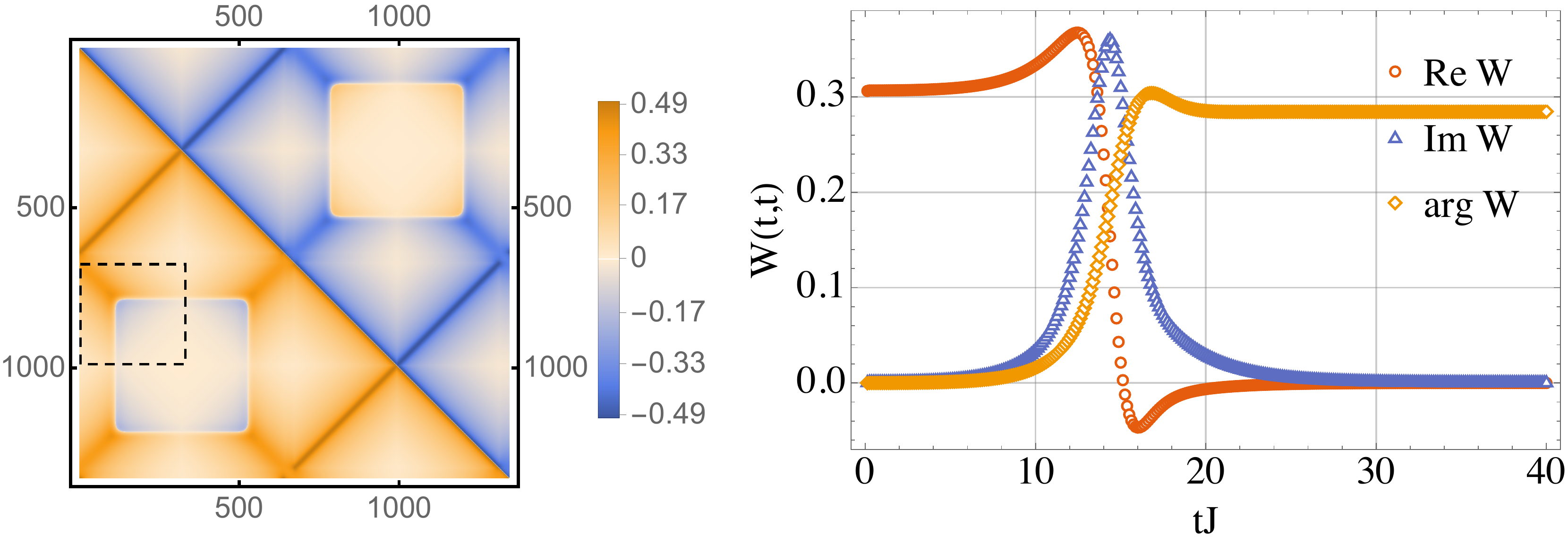}} \\
    \subfloat[The numerical result for $s=0.001e^{-0.95\pi i}$.]{\includegraphics[scale=0.4]{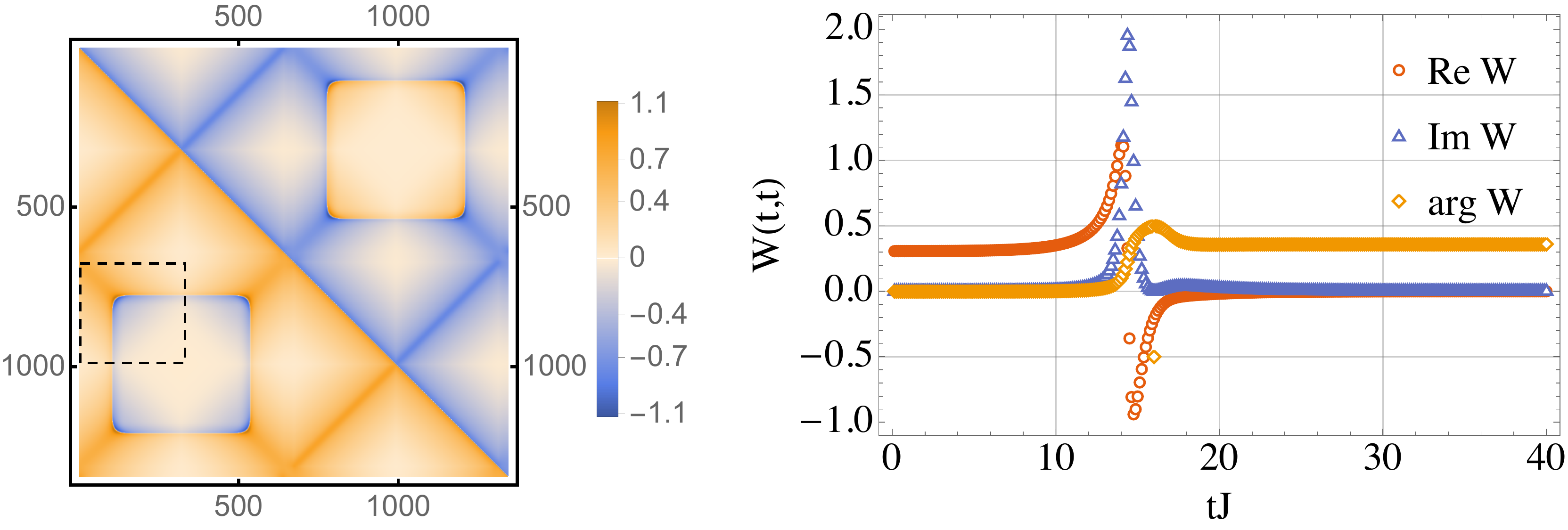}}
  \caption{Numerical results for the $q=4$ model with $\beta J=8$ and $tJ=40$. The color plots on the left show $\Re G(l,l')$; the area representing $W$ is marked by the dashed box. On the right, $\arg W$ is plotted in units of $2\pi$.}
  \label{fignumq4}
 \end{figure}
 
 \begin{figure}[p]
  \center
  \subfloat[The numerical result for $s=0.001$.]{\includegraphics[scale=0.4]{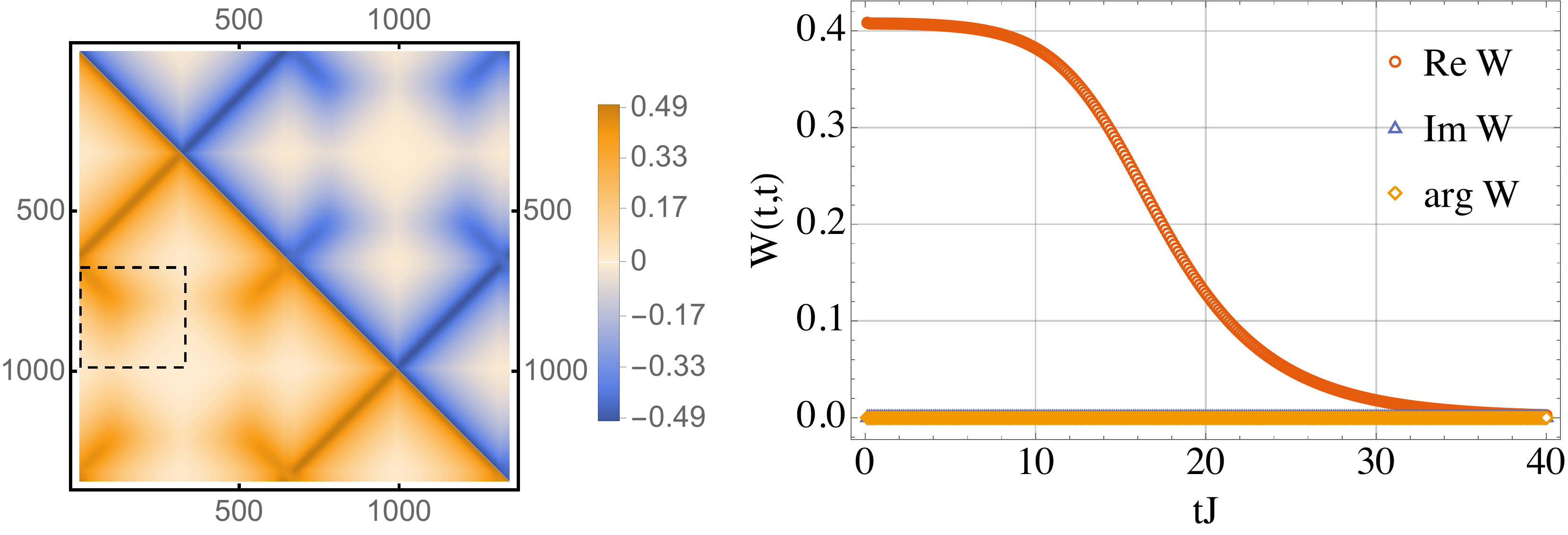}}\\
    \subfloat[The numerical result for $s=0.001e^{-\frac{\pi}{2} i}$.]{\includegraphics[scale=0.4]{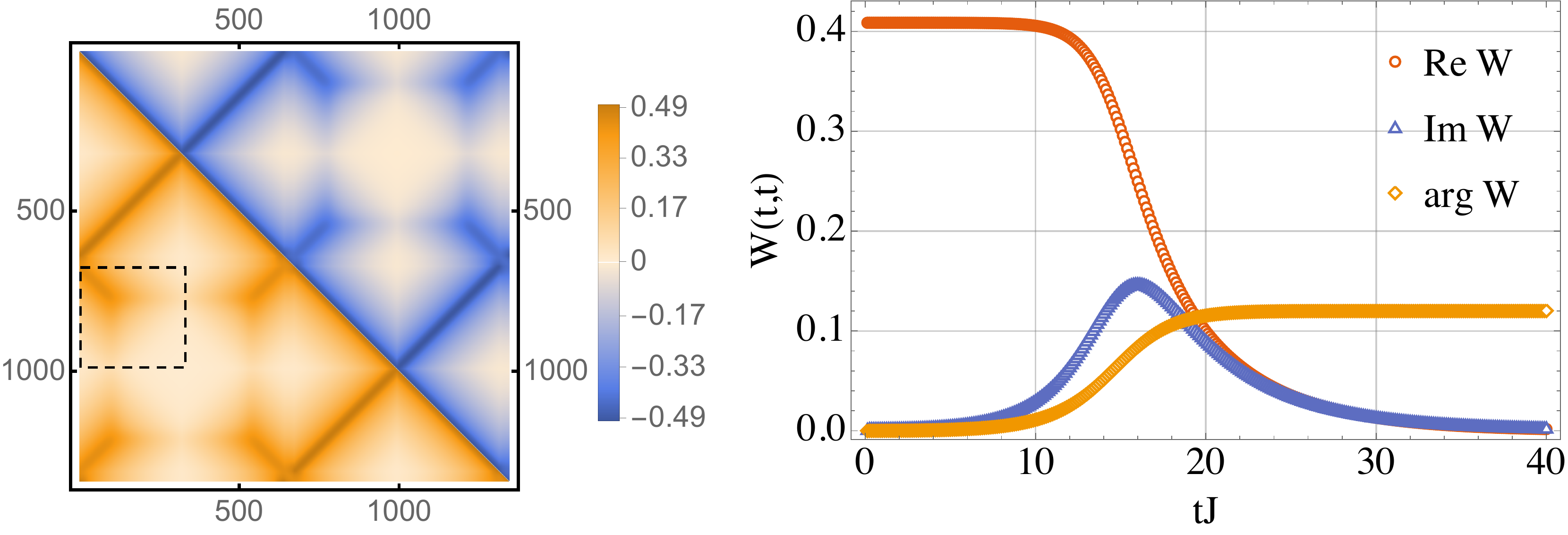}} \\
    \subfloat[The numerical result for $s=0.001e^{-\frac{3\pi}{4} i}$.]{\includegraphics[scale=0.4]{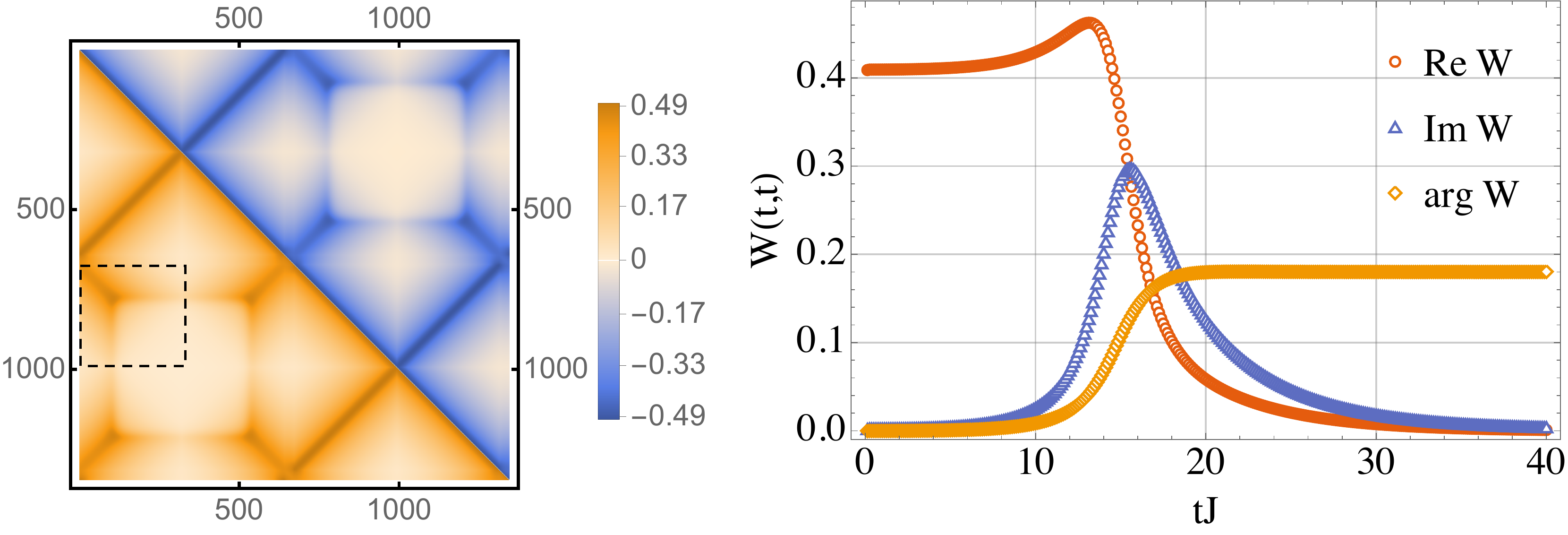}} \\
    \subfloat[The numerical result for $s=0.001e^{-0.95\pi i}$.]{\includegraphics[scale=0.4]{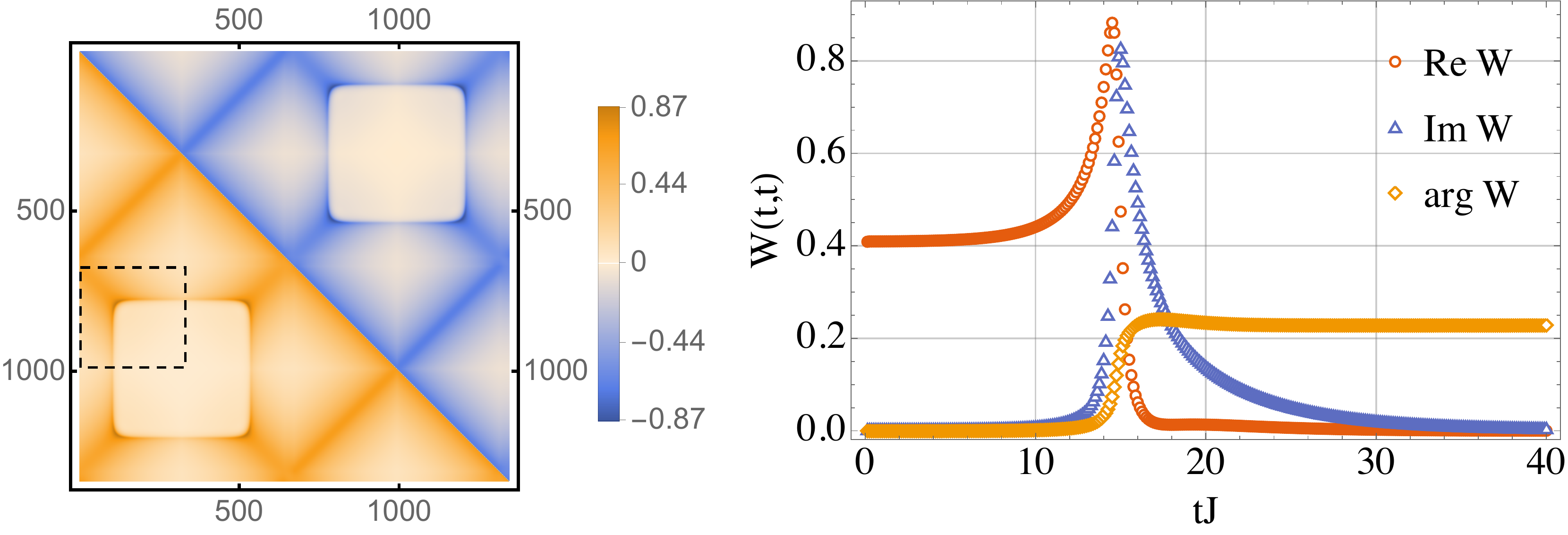}}
  \caption{Numerical results for the $q=6$ model with $\beta J=8$ and $tJ=40$. The color plots on the left show $\Re G(l,l')$; the area representing $W$ is marked by the dashed box. On the right, $\arg W$ is plotted in units of $2\pi$.}
  \label{fignumq6}
 \end{figure}

More detailed results for $q=4$ and $q=6$ are shown in figures \ref{fignumq4} and~\ref{fignumq6}, respectively. In these calculations, we choose $\beta J=8$ and $t J=40$. We take $\Delta l\kern1pt J=1/8$, which corresponds to $L=1344$. When scanning the plane of $s$, we start from $\Re s>0$, where the Green function converges quickly. We then decrease $\Re s$ at a fixed $\Im s$, with the initial guess of $G$ being the last convergent result.

If $s$ is real and positive, $W(t_1,t_2)$ decreases monotonically to zero for both $q=4$ and $q=6$. For $q=4$, when $\arg s$ is large, we find that $\Re W(t,t)$ is non-monotonic and changes sign at some time $t_p\sim \log |s|$ with a finite slope, while $\Im W(t,t)$ shows a peak near $t_p$. However, for $q=6$, there is no sign change. When $|\arg s|\rightarrow \pi$, the slope increases rapidly. For $|\arg s|=\pi$, we find no convergent Green function, implying the divergence of $W(t_1,t_2)$ at some values of $t_1$, $t_2$.

Let us examine the divergence of $W$ at $s<0$ more closely. We may assume that
\begin{equation}
W(t_1,t_2) \propto \bigl(f(t_1-t_2)+se^{\kap(t_1+t_2)/2}\bigr)^{-\alpha}
\end{equation}
near the singularity, which simplifies to $(t_1+t_2-2t_*)^{-\alpha}$ if $t_1\approx t_2$. Then the right-hand side of~\eqref{SD1} diverges as $(t_1+t_2-2t_*)^{-(q-1)\alpha+2}$, where the $2$ in the exponent comes from the integration over $t$, $t'$. This gives the equation $-(q-1)\alpha+2=-\alpha$; hence, $\alpha=2/(q-2)$. By the same hypothesis, if $s$ contains a small imaginary part, namely, if $\arg s=-\pi+\epsilon$, then $\arg W$ jumps by $\alpha\pi$ when $t_1$ or $t_2$ is tuned through the singular point. For $q=4$ and $q=6$, we have $\alpha=1$ and $\alpha=1/2$, respectively. This matches the numerical results shown in figures \ref{fignumq4}\,(d) and \ref{fignumq6}\,(d).

\subsection{High-frequency noise vs.\ Lyapunov exponent}\label{sec:num_tau}

In the main text, we have derived inequality~\eqref{spfun_bound}, which connects the exponential decay rate $\tau_*$ of the spectral function to the Lyapunov exponent $\kap$. The numerical results in the last subsection for $q=4,6$ indicate that $F^\R(z;\frac{\pi}{2}+it_1,-\frac{\pi}{2}+it_2)$ is analytic in $z$ in the entire plane with a branch cut from $-\infty$ to $0$. One possible reason for this is that the bound~\eqref{spfun_bound} might be tight. In this subsection, we numerically compare $\tau_*$ and $\frac{\pi}{\kap}-\frac{1}{2T}$ for the $q=4$ SYK in a broad range of temperatures. 

To find $\tau_*$, we first numerically determine the spectral function $A(\omega)$ using the Schwinger-Dyson equation on the (single) Keldysh contour. At thermal equilibrium, the functions $G^\K(t,t')$ and $G^\R(t,t')$ depend only on $t-t'$; hence, 
\begin{equation}\label{numspfun1}
\tilde G^\R(\omega)=\frac{1}{\omega-\tilde \Sigma^\R(\omega)},\qquad
i\Sigma^{\R}(t) = J^2
\Bigl(\bigl(iG^{\K+\R}(t)/2\bigr)^{q-1}
-\bigl(iG^{\K-\R}(t)/2\bigr)^{q-1}\Bigr),
\end{equation}
where the second equation is a special case of~\eqref{Sigma_Keldysh}. The retarded Green function $G^\R$ is used to compute the spectral function $A$ and then the Keldysh Green function $G^{\K}$:
\begin{equation}\label{numspfun2}
A(\omega)=-2\Im\tilde{G}^\R(\omega),\qquad
\tilde{G}^\K(\omega)=-iA(\omega)\tanh\frac{\beta \omega}{2}.
\end{equation}
Equations~\eqref{numspfun1} and~\eqref{numspfun2} are solved iteratively with discretized $t$ and $\omega$. The parameter $\tau_*$ is obtained from the best exponential fit to the tails of the spectral function.

We then compute the Lyapunov exponent $\varkappa$ by solving the linearized kinetic equation \cite{MS16-remarks,GuKi18}:
\begin{equation}\label{numkappa1}
\left|  \tilde G^{\R}\Bigl(\omega +  i \frac{\varkappa}{2}\Bigr)  \right|^2 \int  \tilde R(\omega-\omega') \tVF^{\R}(\omega') \frac{d\omega'}{2\pi} =\tVF^{\R} (\omega) \,,\quad \quad R(t)=(q-1)J^2\bigl(W^{(0)}(t,0)\bigr)^{q-2}.
\end{equation}  
Here we have introduced $\tVF^{\R}(\omega)=\int_{-\infty}^\infty dt~\VF^\R(\pi+it)$. In numerics, the left-hand side of~\eqref{numkappa1} becomes the vector $\tVF^{\R}$ multiplied by a matrix $K(\varkappa)$. The Lyapunov exponent $\varkappa\in (0,2\pi/\beta)$ is determined by requiring that the largest eigenvalue of $K(\varkappa)$ is $1$.

\begin{figure}[t]\centering
\begin{tabular}{c@{\hspace{1.5cm}}c}
\includegraphics[height=5cm]{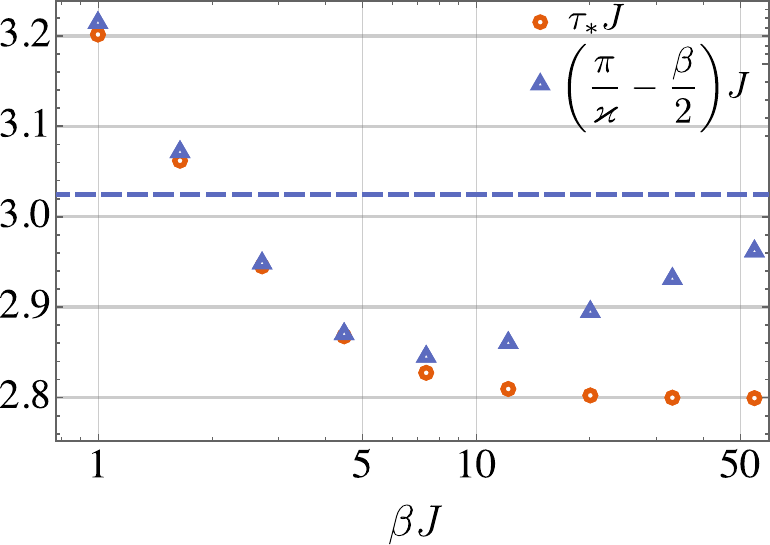} & \includegraphics[height=5cm]{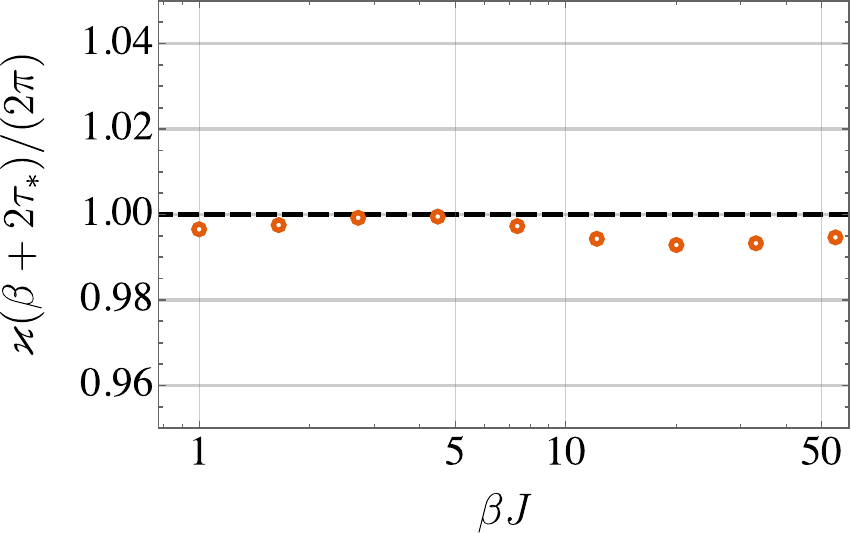}
\vspace{3pt}\\
(a) & (b)
\end{tabular}
\caption{(a) Plots of $\tau_*$ and $\frac{\pi}{\kap}-\frac{\beta}{2}$ as functions of the inverse temperature for the $q=4$ SYK model. The dashed line indicates the $\beta\to\infty$ limit of the second function. (b) The quantity $\kap(\beta+2\tau_*)/(2\pi)$ (or $\kap\bigl(1+\frac{\tau_*}{\pi}\bigl)$ in dimensionless units) that appears in the analyticity domain bound~\eqref{argz_bound}.}
\label{fig:tau_angle}
\end{figure}

The numerical results, plotted in figure~\ref{fig:tau_angle}, are consistent with the inequality $\tau_*\leq \frac{\pi}{\kap}-\frac{1}{2T}$ but show that it is not tight. The $\beta\to\infty$ limit of $\frac{\pi}{\kap}-\frac{\beta}{2}$ is obtained from Eq.~(3.130) (or (3.167) in the arXiv version) of~\cite{MS16-remarks}.

\bibliography{LateTime}
\bibliographystyle{JHEP}

\end{document}